\documentclass[aps,prd,amsmath,amssymb,nofootinbib,showpacs,reprint]{revtex4-1}
\usepackage{pstricks}

\usepackage{etex}
\usepackage[utf8]{inputenc}
\usepackage{color}
\usepackage{graphicx}
\usepackage{bm}
\usepackage{bbm}
\usepackage{pst-grad}

\usepackage{amsfonts}
\usepackage{amsmath}
\usepackage{amssymb}
\usepackage{amsthm}

\usepackage{psfrag}
\usepackage{pstricks,pst-grad,color,shadow}
\usepackage{dsfont}
\usepackage{tikz}
\usetikzlibrary{shapes,arrows,backgrounds}
\usepackage{soul}
\usepackage{placeins}
\usepackage{wrapfig}

\usepackage{pst-all}

\definecolor{darkgreen}{rgb}{0,0.73,0}

\newcommand{\erw}[1]{\left \langle #1 \right \rangle}

\newcommand{\abs}[1]{\left \vert #1 \right \vert}

\newcommand{\algebra}[1]{\mathfrak{#1}}

\newcommand{\gO}{\mathcal{O}}

\newcommand{\trnsp}{\mathsf{T}}

\newcommand{\ft}[2]{{\textstyle\frac{#1}{#2}}}
\DeclareMathOperator{\tr}{tr}

\newcommand{\ii}{\mathrm{i}}
\newcommand{\ON}{O(N) }
\newcommand{\OD}{O(3) }
\newcommand{\SON}{SO(N) }
\newcommand{\pmu}{\partial_{\mu}}
\newcommand{\pnu}{\partial_{\nu}}
\newcommand{\pMu}{\partial^{\mu}}

\newcommand{\Pmu}{\nabla_{\mu}}
\newcommand{\Pnu}{\nabla_{\nu}}

\def\bphi{\boldsymbol{\phi}}

\newcommand{\includeEPSTEX}[1]{\includegraphics{#1}}

\graphicspath{{./}{./publPlots/}}

\begin{document}
\title{Asymptotic safety on the lattice: The Nonlinear \ON Sigma Model}

\newcommand{\FSU}{Theoretisch-Physikalisches Institut, 
Friedrich-Schiller-Universit{\"a}t Jena, 
07743 Jena, Germany}
\newcommand{\JLU}{Institut f\"ur Theoretische Physik, 
Justus-Liebig-Universit\"at Giessen, 35392 Giessen, Germany}

\author{Bj\"orn H. Wellegehausen}
\email{Bjoern.Wellegehausen@uni-jena.de}
\affiliation{\JLU}
\affiliation{\FSU}

\author{Daniel K\"orner}
\email{Daniel.Koerner@uni-jena.de}
\affiliation{\FSU}

\author{Andreas Wipf}
\email{Wipf@tpi.uni-jena.de}
\affiliation{\FSU}


\begin{abstract}
\noindent We study the non-perturbative renormalization group flow of 
the nonlinear \ON sigma model in two and three spacetime dimensions 
using a scheme that combines an effective local Hybrid Monte Carlo 
update routine, blockspin transformations and a Monte Carlo demon 
method. In two dimensions our results verify perturbative 
renormalizability. In three dimensions, we determine the flow diagram 
of the theory for various $N$ and 
different truncations and find a non-trivial fixed point, which 
indicates non-perturbative renormalizability. It is related to the 
well-studied phase transition of the \ON universality class and
characterizes the continuum physics of the model. We compare the 
obtained renormalization group flows with recent investigations by 
means of the Functional Renormalization Group.
\end{abstract}


\maketitle


\section{Introduction}

\noindent 
The renormalization of coupling parameters due to quantum 
fluctuations is a characteristic feature of any quantum field theory
and many different methods have been developed to study this
interesting property. While most of these methods rely on a perturbative
treatment of the theories, the investigation of strongly coupled or
strongly correlated systems without small expansion parameter,
like e.g. the theory of strong interaction, requires a 
non-perturbative approach. One non-perturbative and very flexible 
method is the \emph{Functional Renormalization Group} (FRG)
introduced by K. Wilson \cite{Wilson:1974mb}.
In a particularly useful implementation of the functional renormalization 
group, one studies the flow of the effective average action 
$\Gamma_k$ w.r.t. the momentum scale $k$, which interpolates between the 
bare action at the UV-cutoff $\Lambda$, and the full effective 
action in the IR, $\Gamma_{k\rightarrow 0} = \Gamma$ \cite{Wetterich}.
With the help of this powerful non-perturbative approach one can explore 
theories which are non-renormalizable in perturbation theory,
i.e. in the vicinity of a Gau{\ss}ian fixed point,
but are renormalizable in a non-perturbative setting.
In such asymptotically save theories the running of the couplings 
in the UV is controlled by a non-trivial fixed point with a finite 
number of relevant directions.
The most important theory where this so-called \emph{asymptotic safety scenario}
of Weinberg \cite{weinberg:1979,LivingRev} could be realized
is general relativity where at present all results suggest that there exists
a non-trivial UV fixed point 
\cite{PhysRevD.65.065016,Benedetti:2009rx,Litim:2011cp}.
\\
Here we employ an alternative and efficient non-perturbative approach, 
based on numerical simulations, to study global flow diagrams
of field theories. We apply the technique to spot non-trivial fixed 
points and to determine their properties.\\ 
In order to extract the renormalization of the couplings from lattice 
computations, different methods are used to define the running coupling
such as the renormalized correlation functions or the Schwinger 
functional \cite{Luescher:1991}.
In an alternative recent approach one tries to directly integrate out 
momentum shells on the lattice by using Fourier Monte Carlo 
simulation \cite{Troester:2011}. In the present work we make use
of the well-known \emph{Monte Carlo Renormalization Group} method (MCRG) 
\cite{Hasenfratz1984,Hasenfratz1985,Ma,Lang:1985nw}. It is based 
on the idea of blockspin transformations and can be applied
to theories with fermionic or gauge fields \cite{Catterall2011}. 
By applying successive blockspin transformations, real-space RG-transformations 
are performed and a renormalization trajectory is calculated. 
However, since every RG step typically reduces the linear extent
of the lattice by a factor of $b=2$, exponentially large lattices 
are needed in order to obtain sufficiently long trajectories 
that get close enough to the fixed point regime \cite{Bock1996}. 
Even worse, a standard method to determine the effective couplings 
relies on the matching of correlation functions on the initial and 
blocked lattices and requires expensive scanning runs for the parameters 
of the bare action at the largest lattice used \cite{Shenker1980}. 
In order to circumvent these problems we employ the \emph{demon method} 
\cite{Creutz:1984,Hasenbusch1994,Wozar:2008nv} which allows 
us to efficiently compute RG trajectories at a fixed lattice volume.\\
In the present work we apply the MCRG method in combination
with the demon method to calculate the global flow diagram of the
ubiquitous nonlinear \ON sigma models (NLSM) which are of interest
both in condensed matter physics \cite{Campostrini2006} and in particle 
physics \cite{1984AnPhy.158..142G}. Here they serve as
toy models to test and develop RG methods for models of quantum gravity. 
Both classes of theories share relevant properties. Whereas in two dimensions 
the nonlinear \ON models are perturbatively renormalizable and asymptotically 
free this feature is lost in higher dimensions. But then the
small-$\epsilon$ and $1/N$-expansions both point to the
the existence of non-trivial fixed points in these models
\cite{Polyakov,PhysRevLett.36.691,PhysRevD.14.985,IYa1979393}. 
Their existence is further supported by FRG calculations based on a 
one-parameter truncation of the effective action \cite{Codello2009} and 
higher-order truncations \cite{Flore2012} and we will compare our 
computations with these more recent results.\\
The article is structured as follows: In Sec.~\ref{generalON} we 
discuss general properties of nonlinear \ON models and in Sec.~\ref{MCRG} 
we describe both the MCRG and the demon method. We carefully discuss the 
truncation of the effective action and the optimization of the MCRG method.
In Sec.~\ref{2dresults} we apply the method to the asymptotically free 
two-dimensional sigma model and recover the expected flow
of couplings and fixed point structure.
In Sec.~\ref{3dresults} we study the flow diagram of the three 
dimensional \OD model. We begin with a simple one-parameter truncation 
and then include operators of higher order in the derivatives. 
We also compute the critical exponents and compare the obtained values
with known results. In Sec.~\ref{largeNResults} we continue with
the flow diagrams and critical exponents of \ON models for different 
values of $N$ and study the large-$N$ limit. Our general 
conclusion is contained in Sec.~\ref{Con}. Preliminary
results of this work have been reported in the proceedings \cite{Koerner:2013vxa}.

\section{The \ON nonlinear sigma model in $d$ dimensions} 
\label{generalON}
 
\noindent We recall the Euclidean action of the nonlinear \ON
model with the sphere as target space,
\begin{equation}
S_\sigma=\frac{1}{2g^2}\int d^dx\;\partial_\mu\bphi\cdot\partial^\mu\bphi,
\end{equation}
where $\bphi$ is a N-component scalar field that satisfies the constraint
$\bphi\cdot\bphi=1$. The coupling $g$ has mass dimension
\begin{equation}
 [g]=\frac{2-d}{2}.
\end{equation}
In two spacetime dimensions the global \ON symmetry cannot be broken.
At strong coupling the theory is asymptotically free and the RG flow 
is dominated by a fixed point at infinite coupling, which 
corresponds to a Gau{\ss}ian fixed point for the inverse coupling. Thus,
the model is perturbatively renormalizable. This is not surprising
since in two dimensions the coupling is dimensionless. 
In higher dimensions the coupling has negative mass dimension 
and perturbative renormalizability is lost. However, 
lattice simulations with the discretized action
\begin{equation}
S=\frac{1}{2g^2}\sum_{x,\mu}\;\bphi_x\bphi_{x+\hat\mu}
\end{equation}
reveal a critical point that separates a \ON symmetric phase from a 
broken phase by a second-order phase transition. In the broken phase
there are $N-1$ Goldstone bosons corresponding to the
directions tangential to a sphere in target space.
In order to recover the continuum field theory one may 
use this critical behavior to define the continuum 
limit of the discrete lattice model. 
Much effort went into studying the properties of the model 
near criticality and in particular in calculating its
critical exponents. Thus, a large number of 
results are available, both from numerical high-precision Monte Carlo 
methods as well as analytical calculations using the high-temperature 
expansion or renormalization group method. Even experimental 
data from condensed matter physics are available, see for example
\cite{Shenker1980,Codello2009,Butera1997,Bock1996,
Campostrini2002,Antonenko1998,Hasenfratz1985,Chen1993,Campostrini2006}.\\
We are particularly interested in the flow diagram of the three-dimensional
model that is conjectured to show a non-trivial UV fixed point, 
a necessary requirement for the asymptotic safety scenario to be at work.

\section{Monte Carlo Renormalization group}
\label{MCRG}

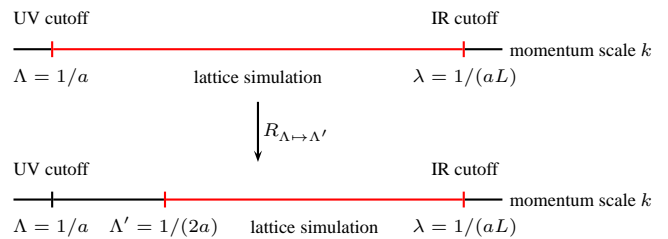
\begin{figure}[h]
\psset{xunit=1cm,yunit=1cm,runit=1cm}
\begin{center}
\begin{pspicture}(0,0)(8,3.5)
 \rput[l](6.6,0.5){\rnode{A}{\scriptsize{momentum scale $k$}}}
 \rput[t](6,1.0){\rnode{B}{\scriptsize{IR cutoff}}}
 \rput[b](6,0.0){\rnode{C}{\scriptsize{$\lambda=1/(a L)$}}}
 \rput[t](0.5,1.0){\rnode{D}{\scriptsize{UV cutoff}}}
 \rput[b](0.5,0.0){\rnode{E}{\scriptsize{$\Lambda=1/a$}}}
 \rput[b](2.0,0.0){\rnode{F}{\scriptsize{$\Lambda^\prime=1/(2a)$}}}
 \rput[b](4.0,0.05){\rnode{G}{\scriptsize{lattice simulation}}}
\psline{-}(0,0.5)(0.5,0.5)
\psline{|-}(0.5,0.5)(2.0,0.5)
\psline[linecolor=red]{|-|}(2.0,0.5)(6,0.5)
\psline{-}(6,0.5)(6.5,0.5)
\rput[l](6.6,2.5){\rnode{A1}{\scriptsize{momentum scale $k$}}}
 \rput[t](6,3.0){\rnode{B1}{\scriptsize{IR cutoff}}}
 \rput[b](6,2.0){\rnode{C1}{\scriptsize{$\lambda=1/(a L)$}}}
 \rput[t](0.5,3.0){\rnode{D1}{\scriptsize{UV cutoff}}}
 \rput[b](0.5,2.0){\rnode{E1}{\scriptsize{$\Lambda=1/a$}}}
 \rput[b](3.25,2.05){\rnode{F1}{\scriptsize{lattice simulation}}}
\psline{-}(0,2.5)(0.5,2.5)
\psline[linecolor=red]{|-|}(0.5,2.5)(6,2.5)
\psline{-}(6,2.5)(6.5,2.5)
\psline{->}(3.25,1.8)(3.25,1.0)
\rput[l](3.25,1.4){\rnode{H}{\scriptsize{ $R_{\Lambda \mapsto \Lambda^\prime}$}}}
\end{pspicture}
\end{center}
\caption{Sketch of the MCRG method}
\label{fig:SketchMCRG}
\end{figure}

\begin{figure*}[tb]
\includeEPSTEX{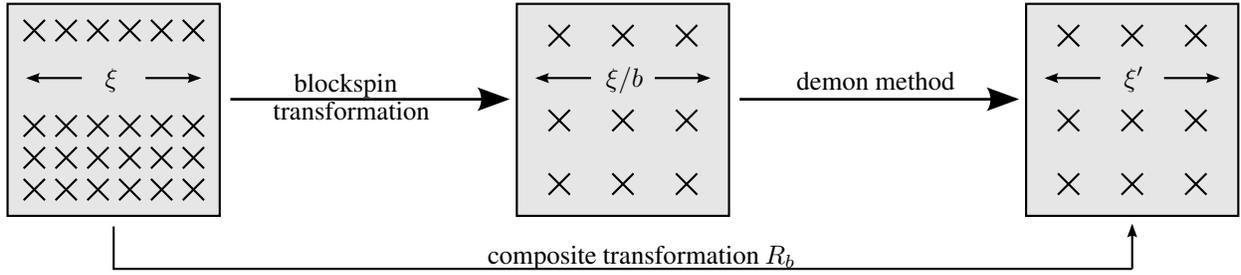}
\caption{The composite transformation $R_{b}$ relates coupling constants
on a lattice with correlation length $\xi$ to a coarser lattice with 
correlation length $\xi'$, which is obtained by using a blockspin 
transformation and mapping the resulting configurations onto a truncated 
effective action by the demon method. Simulating the truncated ensemble 
may not necessarily yield $\xi^{\prime}=\xi/b$ due to truncation errors.}
\label{fig:rgtrafo}
\end{figure*}

\noindent We will study the \ON lattice model at 
zero temperature, i.e. on a lattice with equal temporal and spatial extent 
$L$. The physical volume is hence $V=L^d_\mathrm{phys}=(aL)^d$, 
where $a$ denotes the lattice spacing. In 
Monte Carlo simulations a UV-cutoff at an energy $\Lambda=\pi/a$ 
is introduced naturally and the lattice size $L_\mathrm{phys}$ serves as
IR-cutoff at a lower energy $\lambda=\pi/L_\mathrm{phys}$. On the lattice one may
calculate the $n$-point functions
\begin{equation}
\erw{\phi_{x_1} \ldots \phi_{x_n}}=\frac{\int \mathcal{D}\phi  \,\phi_{x_1}
\ldots \phi_{x_n}\,\exp(-\Gamma_\Lambda[\phi])}{\int \mathcal{D}\,\phi
\exp(-\Gamma_\Lambda[\phi])}
\end{equation}
from which one extracts all physical quantities like e.g. particle masses.
Thereby all quantum fluctuations with scales between the upper and 
lower cutoff are taken into account. The physics at the IR-cutoff
is fixed by choosing a lattice extent $L$ and coupling constants ${g_i}$ 
of the microscopic (bare) action at the UV-cutoff. An RG transformation 
$R_{\Lambda \mapsto \Lambda^\prime}$ relates the parameter 
set $\{g_i\}$ at the high energy scale $\Lambda$ to a parameter set 
$\{g_i^\prime\}$ at lower energy scale  $\Lambda^\prime$,
\begin{equation}
\{g_i\}(\Lambda) \longmapsto \{g_i^\prime\}(\Lambda^\prime)
=R_{\Lambda \mapsto \Lambda^\prime} \left(\{g_i\}\right).
\end{equation}
Thereby the physics, i.e. the $n$-point functions at the lower cutoff 
$\lambda$, remain unchanged. An important property of any such RG 
transformation is that it does not depend on the details of the 
flow in coupling space. In particular the transformation must obey 
the semigroup properties
\begin{equation}
 R_{\Lambda \mapsto \Lambda^\prime}=R_{\Lambda \mapsto \Lambda^{\prime\prime}} 
\circ R_{\Lambda^{\prime\prime} \mapsto \Lambda^\prime}, 
\quad R_{\Lambda\mapsto\Lambda}=1,
\end{equation}
where $\Lambda>\Lambda''>\Lambda'$. This is  depicted in Fig.~\ref{fig:SketchMCRG}.
The infinitesimal change of the couplings is described by the $\beta$-functions
\begin{equation}
 \beta_{i}\left(g\right)=\partial_t g_i\,,\quad t=\ln{\Lambda}.
\end{equation}
The critical exponents $\{\theta_i\}$ of the theory are defined as the negative
eigenvalues of the stability matrix
\begin{equation}
  S_{ij}=\left. \frac{\partial \beta_i}{\partial g_j} \right\vert_{g=g^*}
\end{equation}
at the fixed points $g^*$ of the theory defined by 
$\beta_{i}\left(g^*\right)=0$. Positive critical exponents belong
to relevant direction, negative exponents to irrelevant
directions and vanishing exponents to marginal directions, i.e.
\begin{equation}
 \begin{aligned}
  \theta_i>0 \quad \quad & \text{\emph{relevant} direction}\,,\\
  \theta_i<0 \quad \quad & \text{\emph{irrelevant} direction}\,, \\
  \theta_i=0 \quad \quad & \text{\emph{marginal} direction}.
 \end{aligned}
\end{equation}
By comparing with the scaling of singular thermodynamic observables
near a critical point, one obtains relations between the 
thermodynamic critical exponents and the eigenvalues of the stability 
matrix, for example $\nu=\theta_r^{-1}$ for the critical exponent of the
correlation length $\nu$ and the eigenvalue $\theta_r$ of the 
related relevant direction. \\
In our setup, an RG transformation consists of the two 
steps illustrated in Fig.~\ref{fig:rgtrafo}:
\begin{enumerate}
 \item A blockspin transformation applied to an ensemble with fixed 
	couplings $\{g_i\}$. For the blockspin transformation the 
        semigroup properties are fulfilled.
 \item The demon method to \emph{measure} the effective couplings 
	$\{g_i^\prime\}$ on the blocked lattice. Since this method can only 
	be applied to a truncated effective action the semigroup property 
	of the composite transformation is violated in this step.
\end{enumerate}
In the following we will discuss both steps in more detail.

\subsection{Blockspin transformation}
\noindent
A blockspin transformation with scale parameter $b$ relates 
a field configuration $\{\phi_x\}$ on the fine lattice $(N,a)$ to 
an averaged configuration $\{\phi_x^\prime\}$ on 
the coarser lattice $(N^\prime=N/b,a^\prime=ba)$
\cite{Kadanoff}. The IR-cutoff does not change and the blocked and 
initial configurations describe the same macroscopic physics.
In contrast, the UV-cutoff $\Lambda\rightarrow\Lambda^\prime=\Lambda/b$ 
is lowered and the effective parameters $\{g_i^\prime\}$ 
defined at the new cutoff $\Lambda^\prime$ incorporate the effects of 
all quantum fluctuations with scales between $\Lambda$ 
and $\Lambda^\prime$. A numerical 
simulation on the coarse lattice with couplings $\{g_i^\prime\}$ 
yields the same distribution of averaged fields $\{\phi_x^\prime\}$ as 
obtained  from a simulation at the fine lattice with $\{g_i\}$. 
Each set of parameters defines a point in theory space and they are
connected by an RG trajectory.\\
Here we employ a blockspin transformation where one
draws the averaged fields according to a normalized 
probability distribution,
\begin{equation}
\mathcal{P} (\phi_{x}^{\prime}) \propto \exp \big( C(g)\,\phi_x^\prime\!\! 
\cdot\! R(\{\phi_x\}) \big),\label{eq:blockspintrafo}
\end{equation}
where $R(\{\phi_x\})=\sum_{y \in \square_x}\phi_y$ is the sum over all 
degrees of freedom within a hypercube of the fine lattice.
In our computations we 
choose the smallest cube of size $b^d=2^d$. 
The positive function $C(g)$ determines how strongly the blocked fields 
may fluctuate away from the original degrees of freedom. We shall use 
a function which minimizes the systematic errors induced by the 
unavoidable truncation of the effective action. A detailed discussion
is found in Sec.~\ref{ch:optim}.

\subsection{The demon method} \label{sec:demon}

\noindent With the microcanonical demon method \cite{Creutz:1984} 
one can calculate the couplings in an effective action $S=\sum_{i}g_{i}S_{i}$
such that the corresponding distribution is close to a given ensemble 
of lattice configurations. Hence, given a partition function 
\begin{equation}
 Z(\beta)=\int \mathcal{D}\omega\exp\{-\beta H(\omega)\}
\end{equation}
one introduces an additional degree of freedom $E_D$, the \emph{demon energy},
with the combined partition function 
\begin{equation}
 Z_\text{D}(\beta)=\int \mathcal{D} \omega \int dE_D\,\exp\{-\beta H(\omega)-\beta E_D\}
\label{canDemon}
\end{equation}
of the canonical demon ensemble.
The expectation value of the demon energy can be calculated in a 
simulation of the microcanonical ensemble,
\begin{equation}
 Z_\text{MCD}=\int \mathcal{D} \omega \int dE_D\, \delta\left(H+E_D-E_0\right)
\label{mcanDemon},
\end{equation}
and is related to the inverse temperature $\beta$,
\begin{equation}
 \erw{E_D}=\erw{E_D}(\beta),
\end{equation}
thus allowing to \emph{measure} the inverse temperature $\beta$ 
corresponding to the combined ensemble. 
This method can be generalized to more than one temperature 
or coupling constant, i.e.
\begin{equation}
 Z_\text{MCD}=\int \mathcal{D} \omega \prod \limits_{i} \int dE_D^i\, 
\delta\left(S^i+E_D^i -E_0^i\right).
\label{eq:comb_system}
\end{equation}
Constraining the demon energy to $E_{D}^{i} \in (-E_{m}^{i},E_{m}^{i})$ 
yields
\begin{equation}
	\erw{E_{D}^{i}}_\text{D} = \frac{1}{g_{i}} - \frac{E_{m}}{\tanh(g_{i} 
	E_{m})}\approx \erw{E_{D}^{i}}_\text{MCD}
\end{equation}
where the subscript D denotes the canonical demon ensemble \eqref{canDemon} 
and MCD the microcanonical demon ensemble \eqref{mcanDemon}.
This equation can be solved by numerical means and is used to extract 
the coupling constants $\{g_{i}\}$ from the mean demon energies 
on the right hand side. In the microcanonical ensemble 
the total energy is fixed. Since we want to measure the couplings 
of the blocked ensemble without interference from the demon, we 
demand that $\abs{E_D^i}\ll \abs{S^i}$. Then the algorithm for our MCRG 
setup reads as follows:
\begin{enumerate}
 \item Pick a configuration distributed according to
      the canonical ensemble with action $S=g_1S_1+g_2S_2+\dots$ on 
      the fine lattice.
 \item Perform a blockspin transformation on this configuration.
 \item Use the result as starting configuration
       for a microcanonical simulation of the combined system
       (\ref{eq:comb_system}) and 
        measure the mean demon energies. The starting values 
        for the demon energies are given by the mean demon energies 
        extracted from the previous microcanonical runs.
 \item Repeat step one to three until a sufficient number of 
	configurations has been generated.
 \item Calculate the couplings $g_i^\prime$ from the mean demon energies.
\end{enumerate}
A comparison of these couplings with the initial 
ones yields an approximation for the running of the coupling. It reads:
\begin{equation}
\beta_{i}\left(g\right)=\partial_t g_i=-a \frac{\partial g_i}{\partial
a}=-(g_i^\prime-g_i).
\label{eq:betafunc}
\end{equation}
Note that the $g_i$ are the dimensionless couplings on the 
lattice. In order to measure the critical exponents, we introduce a 
hypercubic grid in coupling space with spacings $\delta g_i$ and 
compute the matrix $S_{ij}$,
\begin{equation}
S_{ij}(g)=\frac{\partial \beta_i}{\partial g_j}=\frac{
\beta_i(g_j+\delta g_j)-\beta_i(g_j-\delta g_j)}{2\delta g_{j}}\,.\label{eq:stabmatw}
\end{equation}
The additional systematic error from discretizing the second derivative 
can be made arbitrarily small by choosing a finer grid in the space
of couplings. Finally we compute the matrix $S_{ij}(g)$ 
and its eigenvalues $\theta_i(g)$ at couplings in the vicinity of a 
critical point $g^*$ to obtain the critical exponents 
and associated thermodynamical critical exponents.

\subsection{Truncated effective action}

\noindent In general, more and more operators are generated 
by the repeated application of the blockspin transformations
and this yields a trajectory in theory space. Since it is impossible
to keep track of all operators we
restrict our analysis to an  ansatz for the effective action 
that only includes a finite number of operators. Thereby, the demon method 
leads to a projection of RG trajectories from general theory space down to modified 
trajectories in a truncated theory space that only consists of the terms
contained in the effective action. Naturally, this procedure introduces 
additional systematic uncertainties which we will denote as truncation 
errors. A qualitative understanding of the truncation errors is obtained
by comparing different truncations. For this reason, we utilize a systematic 
derivative expansion of the effective action up to fourth order. 
In the continuum formulation it is given by 
\begin{equation}
\label{contAction}
	S[\bphi]=\sum \limits_{i=0}^3 g_i N S_i[\bphi] + \mathcal{O}(\partial^6)
\end{equation}
with operators
\begin{align}
	S_{0} &= -\int d^{d}x\; \bphi\cdot \partial_{\mu}\partial^{\mu}\bphi \\
	S_{1} &= \int d^{d}x\; \bphi\cdot (\partial_{\mu}\partial^{\mu})^2\bphi \\
	S_{2} &= \int d^{d}x\; (\bphi\cdot\partial_{\mu}\partial^{\mu}\bphi)^{2} \\
	S_{3} &= \int d^{d}x\;
	(\bphi\cdot\partial_{\mu}\partial^{\nu}\bphi)(\bphi\cdot\partial^{\mu}
        \partial_{\nu}\bphi).
\end{align}
Note that we have introduced an additional factor $N$ in \eqref{contAction} 
in order to get rid of the leading $N$ dependence  of the couplings $g$. 
The couplings have mass dimension
\begin{equation}
 [g_0]=d-2\,,\quad [g_i]=d-4 \quad \text{for} \quad i=1,2,3.
\end{equation}
This is a complete set of the fourth order operators that are compatible 
with the symmetries of the model. Note that other RG studies (like 
the one in \cite{Flore2012}) employ an alternative 
parametrization of the effective action. The relation between these two 
formulations is explained in Appendix \ref{expRel}.\\
Now we discretize the action \eqref{contAction} on a hypercubic lattice,
\begin{equation}
S\big(\{\phi_x\}\big)=\sum \limits_{i=0}^3 g_i N S_i^\prime\big(\{\phi_x\}\big),
\end{equation}
where a straightforward discretization of the continuum operators is given by
\begin{widetext}
\begin{align}
S_0^\prime= & 2 \sum \limits_{x,\mu} \Phi_x \cdot \Phi_{x+\mu}-2dV \\
S_1^\prime= & 2 \sum \limits_{x,\mu,\nu} \Phi_x \cdot
\left(\Phi_{x+\mu+\nu}+\Phi_{x+\mu-\nu}\right)- 4 d \sum \limits \Phi_x
\cdot \Phi_{x+\mu}+4 d^2 V \\
S_2^\prime= & \sum \limits_{x,\mu,\nu} \left \lbrace \left(\Phi_x \cdot
\Phi_{x+\mu}\right)\left(\Phi_x \cdot \Phi_{x+\nu}\right)+ \left(\Phi_x \cdot 
\Phi_{x-\mu}\right)\left(\Phi_x \cdot
\Phi_{x-\nu}\right)+ 2\left(\Phi_x \cdot\Phi_{x+\mu}\right)\left(\Phi_x \cdot
\Phi_{x-\nu}\right)\right \rbrace - \nonumber\\
&8 d \sum \limits_{x,\mu}\Phi_x \cdot \Phi_{x+\mu}+4 d^2 V \label{eq:ops} \\
S_3^\prime= & \sum \limits_{x,\mu,\nu} \left \lbrace 
\left(\Phi_x \cdot \Phi_{x+\mu}\right)\left(\Phi_x \cdot \Phi_{x+\nu}\right)+
\left(\Phi_x \cdot \Phi_{x-\mu}\right)\left(\Phi_x \cdot \Phi_{x-\nu}\right)-
2 \left(\Phi_x \cdot \Phi_{x+\mu}\right)\left(\Phi_x \cdot
\Phi_{x+\nu-\mu}\right) \right. \nonumber\\ - & \left. 2 \left(\Phi_x \cdot
\Phi_{x-\mu}\right)\left(\Phi_x \cdot \Phi_{x+\mu-\nu}\right)+ \left(\Phi_x 
\cdot \Phi_{x+\nu-\mu}\right)\left(\Phi_x \cdot
\Phi_{x+\mu-\nu}\right) + 2 \left(\Phi_x \cdot \Phi_{x+\mu-\nu}\right)\right
\rbrace \nonumber\\ + & 2d \sum \limits_{x,\mu} \left(\Phi_x \cdot
\Phi_{x+\mu}\right)\left(\Phi_x \cdot \Phi_{x-\mu}\right)-4d \sum
\limits_{x,\mu} \Phi_x \cdot \Phi_{x+\mu}+d^2 V.
\end{align}
\end{widetext}
This set of lattice operators forms a basis of the space of fourth-order 
derivative operators, but it is not orthogonal in operator space. In 
order to improve the convergence of the demon method it is useful to 
reparametrize the action functional in terms of the operators 
$S_i^\prime$ given by
\begin{equation}
\begin{pmatrix} S_0^\prime \\ S_1^\prime \\S_2^\prime \\
S_3^\prime \end{pmatrix} = 
\begin{pmatrix} 2 & 0 & 0 & 0 \\ -4d & 2 & 0 & 0 \\-8d & 0 & 1 & 0 \\ 
-4d & 0 & 0 & 1 \end{pmatrix} \begin{pmatrix} \hat S_0 \\ \hat S_1 \\ \hat S_2 \\ \hat S_3 \end{pmatrix}
 + \begin{pmatrix} -2dV \\ 4d^2V \\ 4d^2V \\ 
d^2 V \end{pmatrix}.
\end{equation}
For simplicity we drop the hat over lattice quantities in the following.

\subsection{Optimized blockspin transformation}\label{ch:optim}

\noindent In a lattice simulation we have access to observables (like 
e.g. the masses) which receive contributions from \emph{all possible} 
lattice operators. This information, which is in part lost if one uses a
truncated effective action, allows us to 
extend our analysis of truncation errors. The macro-physics
is completely determined by the correlation functions and hence must 
agree for the original and blocked ensemble  in Figure \ref{fig:rgtrafo}, 
since the blockspin transformation does not change the IR physics. Applying
the demon method leads to a truncated ensemble which is defined by the
effective action. In general, the correlation functions of the blocked
and truncated ensemble do not coincide. This discrepancy is solely 
due to the truncation of the effective action. In addition to the
simulation of the blocked ensemble, we also simulate the truncated 
ensemble in order to measure the difference in the correlation functions
and thus quantify the systematic truncation errors directly.\\
We reduce this difference by adjusting the blockspin 
transformation. The location of the renormalized trajectory depends on 
the chosen renormalization scheme \cite{Hasenfratz1984} and we aim
at constructing a scheme for which the renormalized trajectory is closest to our truncated 
effective action. More accurately, we have used the improved blockspin transformation 
(\ref{eq:blockspintrafo}) and tuned the  free parameter $C$. In general, 
the optimal value depends on the coupling constants, lattice size, 
target space and number of RG steps. Only in the ideal world
without truncation we expect our results to be independent 
of the RG scheme and thus the optimization constant $C$.\\
In order to tune the constant $C$ in the improved 
blockspin transformation, we compare the correlation lengths extracted from
the two-point-functions on the fine and coarse lattice
and ignore all other correlation functions. Blockspin transformations 
reduce the lattice correlation length $\xi$ exactly by a factor $b$ and
thus we demand the correlation length $\xi'$ in the truncated ensemble to be
equal to $\xi/b$ in order to minimize truncation errors. For simplicity, 
we allow the optimization constant to depend linearly on the couplings,
\begin{equation}
 C(g)=\sum \limits_i c_i g_i\,, \quad c_i=\text{const.}\, . 
\end{equation}
It is clear from the structure of (\ref{eq:blockspintrafo}) that the choice
$C=0$ leads to a complete loss of information and thus results 
in a trivial flow diagram. In most of the 
following computations we find that it is sufficient to tune only the 
first parameter $c_0$ since in the vicinity of the non-Gau{\ss}ian fixed points the 
corresponding couplings $g_i$ are small compared to $g_0$. Nevertheless, 
a small but non-vanishing value for the other $c_i$ is necessary to 
improve the flow in the vicinity of the Gau{\ss}ian fixed point. 
Finally we note that the lattice itself together
with the blockspin transformation acts as the regulator function in 
FRG calculations. Tuning the ratio of correlation functions to the 
optimal ratio corresponds to the choice of an optimal regulator in the 
FRG framework. Roughly speaking it minimizes the \emph{flow time} (RG steps) 
from the UV to the IR.

\section{The RG flow in two dimensions}
\label{2dresults}
\begin{figure}[htb]
\scalebox{1}{\includeEPSTEX{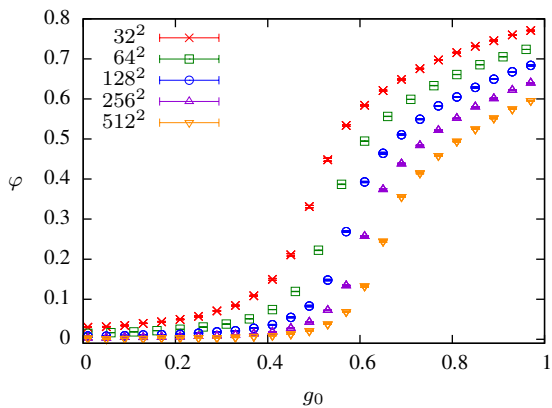}}
\caption{The average field expectation value is shown as a function of 
$g_0$ for different lattice sizes and $N=3$ for $S=g_0 N S_0$.}
\label{fig:2daverageField}
\end{figure}

\begin{figure}[htb]
\scalebox{1}{\includeEPSTEX{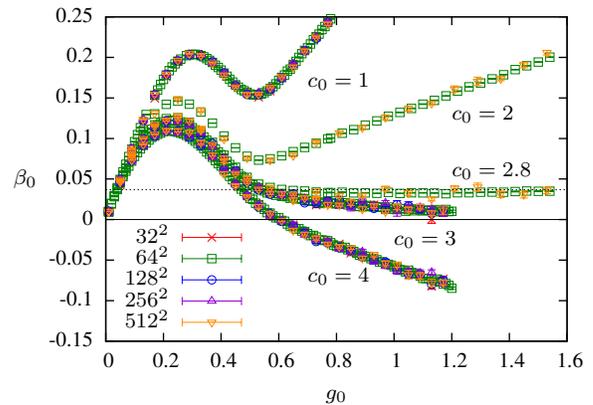}}
\caption{The beta function $\beta_0$ for the simplest possible
truncation and $N=3$ is almost independent of the lattice volume. 
For $c_0<2.8$ it possesses only one fixed point at vanishing coupling. 
For $c_0=2.8$ it becomes constant for $g \to \infty$. The dotted line 
represents the analytical result for $N \to \infty$ and $g \to \infty$. For 
$c_0>2.8$ we find an additional fixed point at finite coupling which is
an artifact of the truncation. }
\label{fig:2dbetaop1}
\end{figure}

\begin{figure}[htb]
\scalebox{1}{\includeEPSTEX{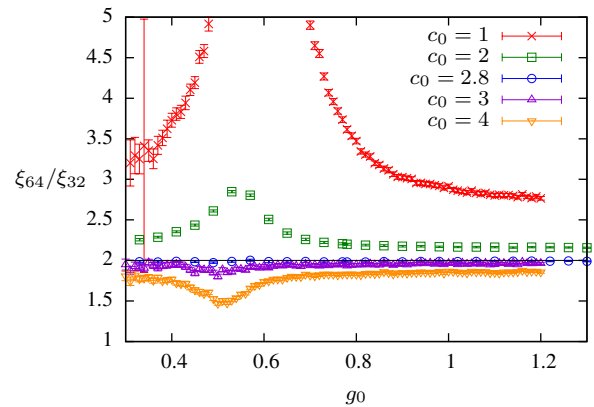}}
\caption{The ratio of the correlation length for a $64^2$ and $32^2$ 
lattice is shown for different parameters of the RG transformation.}
\label{fig:2dmass}
\end{figure}

\begin{figure}[htb]
\scalebox{1.1}{\includeEPSTEX{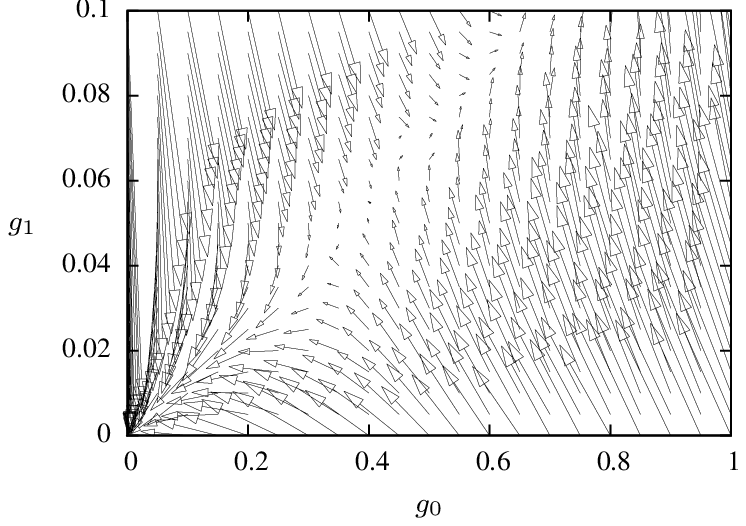}}\\
\scalebox{1.1}{\includeEPSTEX{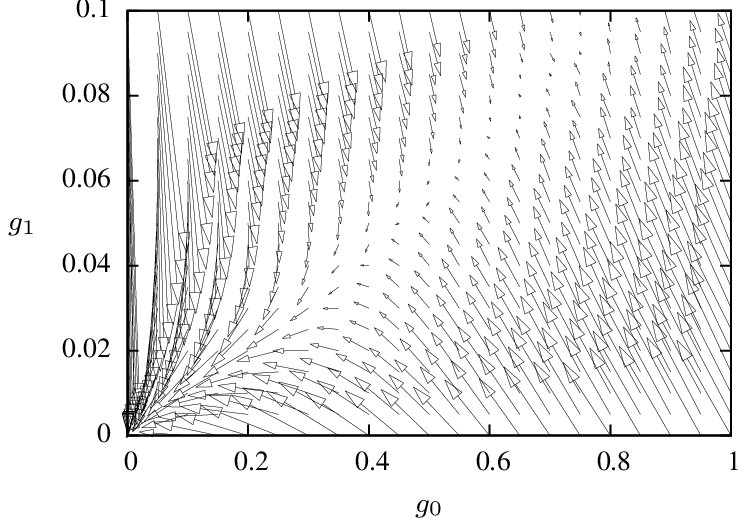}}\\
\scalebox{1.1}{\includeEPSTEX{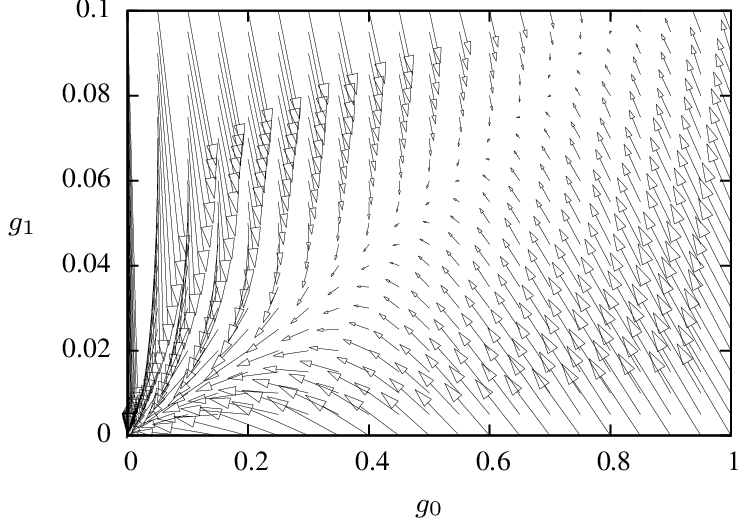}}
\caption{Shown is the flow diagram for $c_0=3$ and $N=3$ in the two operator 
truncation on a $16^2$ (upper panel), $32^2$ (middle panel) and $64^2$ 
lattice (lower panel).}
\label{fig:2d2opflow}
\end{figure}

\noindent In order to test and optimize our method, we reproduce the
beta function for the two dimensional \ON sigma model, which has already 
been computed using the MCRG matching method for $N=3$ \cite{Shenker1980} 
and $N\rightarrow\infty$ \cite{Hirsch1983,Hasenfratz1984}.
The coupling constant $g_0$ of the standard action 
$S_0$ is dimensionless and the theory is thus perturbatively renormalizable. 
From asymptotic freedom we expect that the flow diagram contains two 
trivial fixed points, one in the IR at vanishing coupling and the other 
in the UV at infinite coupling, i.e. vanishing inverse coupling 
\cite{Bock1996}. However for numerical 
simulations only finite lattices are accessible and the theory 
possesses a transition from a symmetric regime at low coupling 
(large physical volume) to an ordered regime at strong coupling (small  physical volume). 
The expectation value of the scalar field,
\begin{equation}
 \varphi=\erw{\left \vert\frac{1}{V}\sum \limits_x \Phi_x\right \vert},
\end{equation}
is shown in Fig.~\ref{fig:2daverageField} as a function of the coupling 
for different lattice sizes.
With increasing volume the transition shifts to larger values 
of the coupling and we conclude that in the infinite volume limit the 
theory is in the symmetric regime for every finite value of the 
coupling, as predicted by the Mermin-Wagner theorem. It is also
evident that finite volume effects are more important for large coupling. 
In particular, the observed behaviour might mimic an additional 
non-trivial fixed point of the RG flow.
In Fig.~\ref{fig:2dbetaop1} we show the $\beta$-function for the 
coupling $g_0$ in the simplest truncation using only the operator 
$S_0$.
We observe that while the $\beta$-function is independent of the lattice 
volume, it depends on the parameter $c_0$ of the RG transformation.
For $c_0=1$ the $\beta$ function has an IR fixed point at vanishing 
coupling and stays positive even for large coupling. Tuning $c_0$ to 
larger values, the $\beta$-function develops a further zero crossing at
finite coupling. However, this additional zero of the $\beta$-function is an
artifact of the truncation. In Fig.~\ref{fig:2dmass} we show 
the ratio of correlation lengths of the original ensemble on the $64^2$ 
lattice compared to the truncated ensemble on the $32^2$ lattice. 
Truncation errors are assumed to be minimal for $\xi_{64}/\xi_{32} = 2$.
For $c_0=1$ and $c_0=4$ significant deviations are visible. We find that
$c_0=2.8$ provides a good matching for a large range of couplings. The 
corresponding beta function in Fig.~\ref{fig:2dbetaop1} does not show an 
additional zero crossing, which coincides with earlier results \cite{Shenker1980}. 
For large $g_0$ the $\beta$ function approaches a constant value corresponding 
to the large $N$ result $\beta(N \to \infty,g \to \infty)= \ln(2)/(6 \pi)$ 
\cite{Hirsch1983}. In order to further improve on our truncation, we add 
a second operator $S_1$ and the resulting flow diagram for fixed $c_0=3$ and 
different lattice sizes is shown in Fig.~\ref{fig:2d2opflow}.
The flow is no longer independent of the volume and for the smallest 
lattice, which is $16^2$, an additional fixed point in the 
($g_0$, $g_1$)-plane emerges. 
However, going to larger lattice volumes, this fixed point shifts away
towards larger couplings and thus we assume that in the continuum 
limit no additional fixed point of the RG flow exists.\\
The renormalized trajectory is the single trajectory that connects the 
Gau{\ss}ian fixed point at the origin with the trivial fixed point 
at infinite coupling. The arrows plotted in Fig.~\ref{fig:2d2opflow} 
point towards the IR and therefore the fixed point at the origin
is an IR fixed point while the fixed point at infinite coupling is UV 
attractive. Again we find that the structure of the flow diagram
using the two-operator truncation matches the prediction from
asymptotic freedom. The known results are very well reproduced with 
our method and we proceed with the \ON models in three spacetime 
dimensions.

\section{Fixed Points of the RG Flow in three dimensions}
\label{3dresults}

\begin{figure}[htb]
\scalebox{1}{\includeEPSTEX{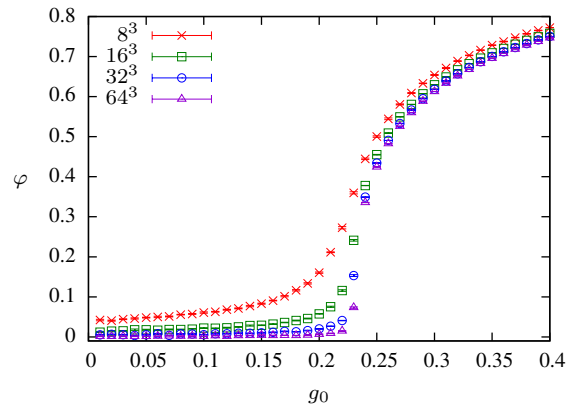}}
\caption{The average field expectation value is shown as a function of 
$g_0$ for different lattice volumes.}
\label{fig:3daverageField}
\end{figure}

\noindent
As in two dimensions we first investigate the \OD model. In 
Fig.~\ref{fig:3daverageField} the order parameter $\varphi$ for the 
spontaneous breaking of the \ON symmetry is shown. 
The critical coupling in the thermodynamic limit is given 
by $g_0^c=0.6862385(20)/3=0.2287462(7)$ \cite{Campostrini2002}. On a $32^3$ lattice, 
lattice artifacts are already sufficiently small for our purpose. 
Therefore most RG transformations considered in the present work are
based on a transformation for a fine lattice with $32^3$ points
to a coarse lattice with $16^3$ points. The critical coupling 
on the $32^3$ lattice is $g_0^c=0.22975(25)$.

\subsection{One-parameter effective action}

\begin{figure}[htb]
\scalebox{1}{\includeEPSTEX{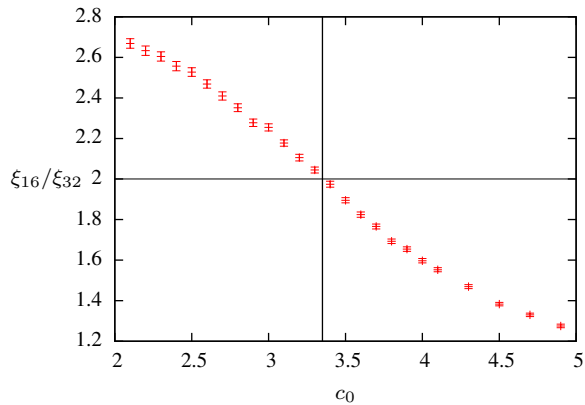}}
\caption{The ratio of correlation lengths obtained by blocking a 
$32^{3}$ lattice down to $16^{3}$ using different optimization 
constants $c_{0}$. A value of $\xi_{16} / \xi_{32}=2$ is 
expected to minimize truncation 
errors and we read off the optimal value $c_0^{\text{opt}}=3.35$ for $N=3$.}
\label{fig:trunctest}
\end{figure}

\begin{figure}[htb]
\scalebox{1}{\includeEPSTEX{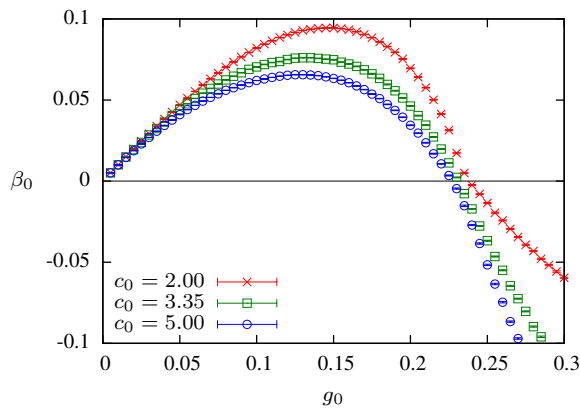}}
\caption{The $\beta$ function for the $1\rightarrow 1$ truncation in
three dimensions and $N=3$ is shown for different values of $c_0$.}
\label{fig:beta11}
\end{figure}

\noindent 
We begin with the simplest truncation possible by using
the one-parameter action $S=g_0 N S_0$. 
We denote this scheme as $1\rightarrow 1$ truncation, indicating the 
use of the one-parameter
action in both ensemble creation and effective action ansatz.\\
As in two dimensions, the $\beta$ function for this truncation is 
almost independent of the lattice size. Using different sizes, we see 
that our results from $8^3$ and $16^3$ already agree within their
statistical error bars and therefore we are confident that our simulations
on a lattice with $32^3$ points do not suffer from large finite size effects.\\
In order to determine the optimization constant in the
blockspin transformation, we again consider the correlation length of 
the two-point function. A perturbative calculation \cite{Hasenfratz1984} 
yields $c_0^\text{pert}=2.3$ for arbitrary N and a large number of 
subsequent RG steps. But computing the ratio of correlation lengths (see
Figure \ref{fig:trunctest}), we see that there exists an optimal choice 
$c_0^{\text{opt}}=3.35$ which leads to the desired value of 
$\xi_{16}/\xi_{32}=2$. This value deviates significantly
from $c_0^\text{pert}=2.3$, indicating that the non-trivial fixed 
point is indeed a non-perturbative feature of the theory.
Already with this simple setup, we find that the dimensionless $\beta$ 
function, depicted in Fig. \ref{fig:beta11}, exhibits the 
qualitative features that were expected from other 
methods \cite{Polyakov,PhysRevLett.36.691,PhysRevD.14.985,IYa1979393,Codello2009}.
In contrast to the two dimensional case, the $\beta$ function 
shows a non-trivial fixed point $g_0^*$ with $\beta(g_0^*)=0$ for every 
value of $c_0$. This clearly points to the non-perturbative 
renormalizability of the \OD-model and is directly related to a 
second-order phase transition. For
the optimal choice $c_0^{\text{opt}}$ we obtain $g_0^*=0.2310(5)$. Systems 
with bare coupling $g_0<g_0^*$ flow to the disordered phase in the IR 
which is controlled by the Gau{\ss}ian fixed point at $g_0=0$, while 
systems with bare coupling $g_0>g_0^*$ flow to the completely ordered 
phase described by $g_0=\infty$ or $1/g_0 = 0$.
These two fixed points correspond to the expected low-temperature fixed point at 
infinite coupling (absolute order) and the expected
high-temperature fixed point at 
zero coupling (absolute disorder). The critical hypersurface is reduced 
to a single point $g_0^*$ in this truncation and the operator $S_0$ 
corresponds to a \emph{relevant} direction of the RG flow.\\ 
Using the information provided by thermodynamical observables like e.g. 
the susceptibility of the order parameter, we can determine the 
\emph{critical point} $g_0^c$ where the correlation length of the system 
diverges at infinite volume. In general theory space, it is the 
point of intersection between the critical hypersurface and the line
where $g_i=0$ except $g_0$. A lattice simulation starting at $g_0^c$ in the UV
will flow along the critical line into the 
non-trivial fixed point and observables measured on this ensemble reflect
the macroscopic physics at this point. Please note that $g_0^c$ need not 
be identical to $g_0^*$ due to truncation errors that affect the value 
for $g_0^*$. Of course, without truncation errors the fixed 
point is located at the critical surface.
We now proceed to discuss higher-order truncations which take additional
operators into account and provide a more complete picture of the flow 
of the effective action. 

\subsection{Higher-order truncations}

\begin{figure}[htb]
\includeEPSTEX{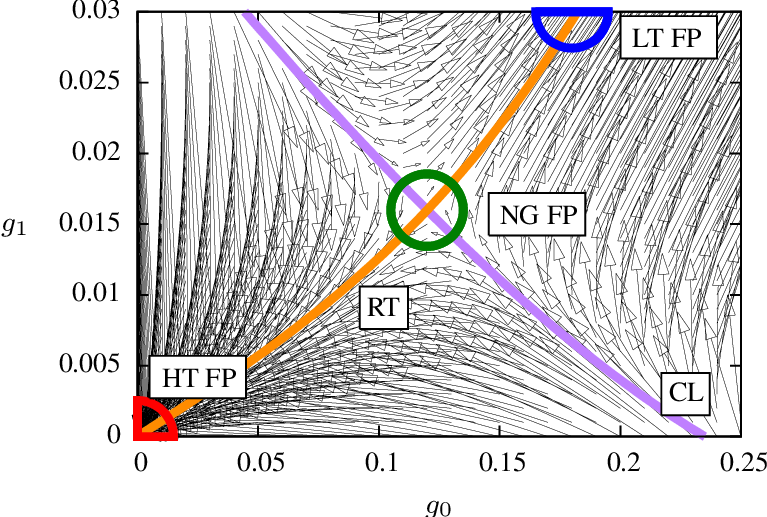}\\
\includeEPSTEX{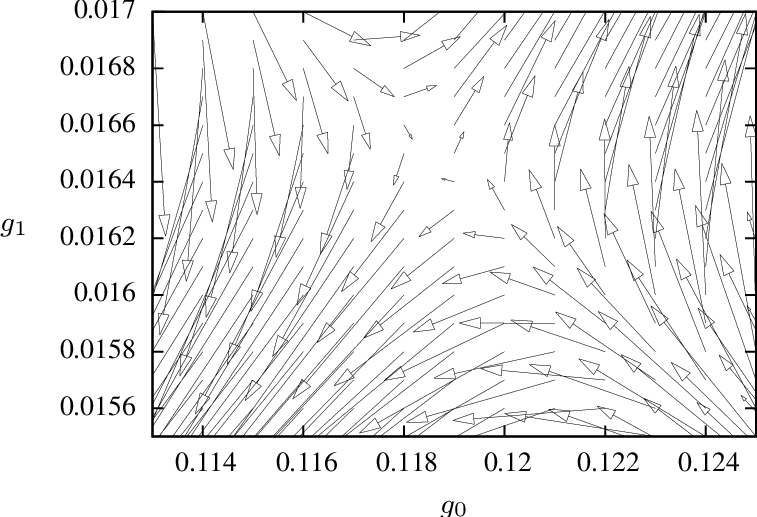}
\caption{The flow diagram using the $2\rightarrow 2$ truncation in three 
dimensions and $N=3$ clearly shows a non-Gau{\ss}ian fixed point (NG FP) 
in the center of the plot in the upper panel. The critical line (CL) and 
renormalized trajectory (RT) intersect at the NG FP. The lower panel shows the 
vicinity of the NG FP. The RG parameters for this flow diagram are $c_0=3.1$ and $c_1=2.5$.}
\label{fig:flowop22}
\end{figure}

\noindent
In the preceding sections we have seen that near the non-trivial fixed point
the operator $S_0$ defines a relevant direction. In this section
we include more operators in the effective action in order to find
the total number of relevant directions. Figure \ref{fig:flowop22} (upper panel) 
shows the global flow diagram for the truncation using two operators 
$S=g_0 N S_0+g_1 N S_1$, both for ensemble generation as well as in the 
demon method ($2\rightarrow 2$ truncation). 
\begin{figure}[htb]
\includeEPSTEX{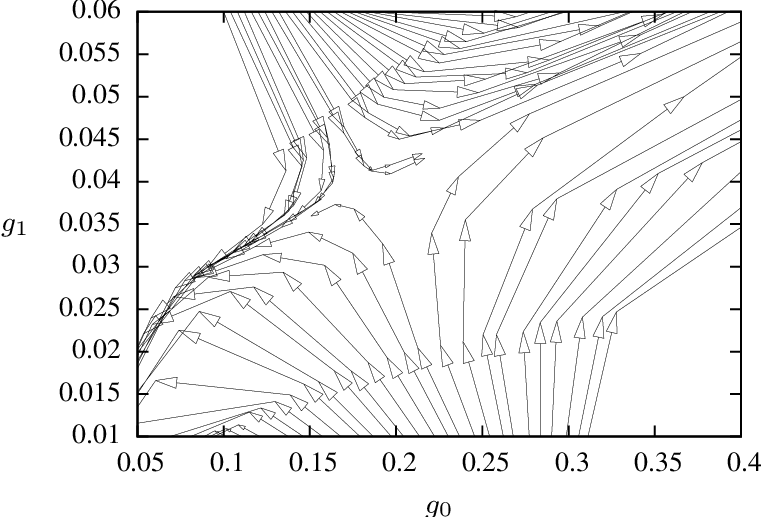}\\
\includeEPSTEX{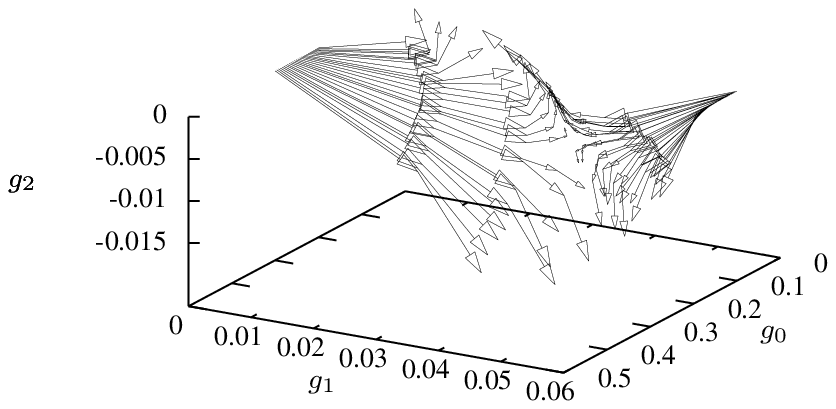}
\caption{Using a shooting technique, the RG trajectories for the 
$3\rightarrow 3$ truncation with operators $S_0,S_1$ and $S_2$
reveal an analogous structure to the 
$2\rightarrow 2$ case. The projection on the $g_0$-$g_1$ axis in the
upper panel shows only a single relevant direction at the non-Gau{\ss}ian 
fixed point. The lower panel shows that the trajectories first approach 
the fixed point regime and afterwards flow along the renormalized 
trajectory to the respective IR fixed points.}
\label{fig:flowop33}
\end{figure}
\begin{figure*}[htb]
\scalebox{0.98}{\includeEPSTEX{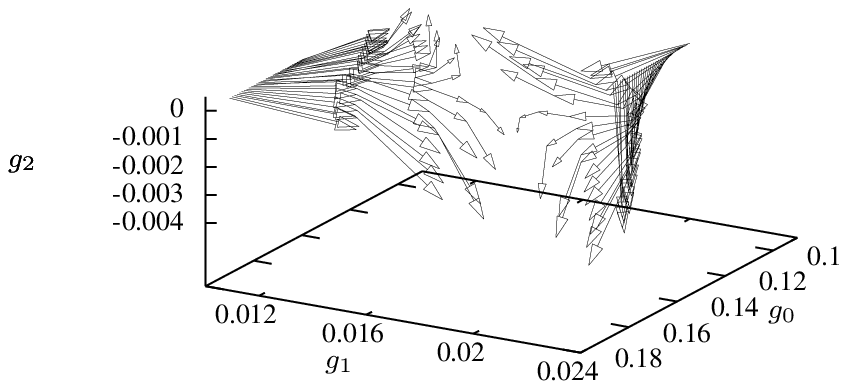}}
\scalebox{0.98}{\includeEPSTEX{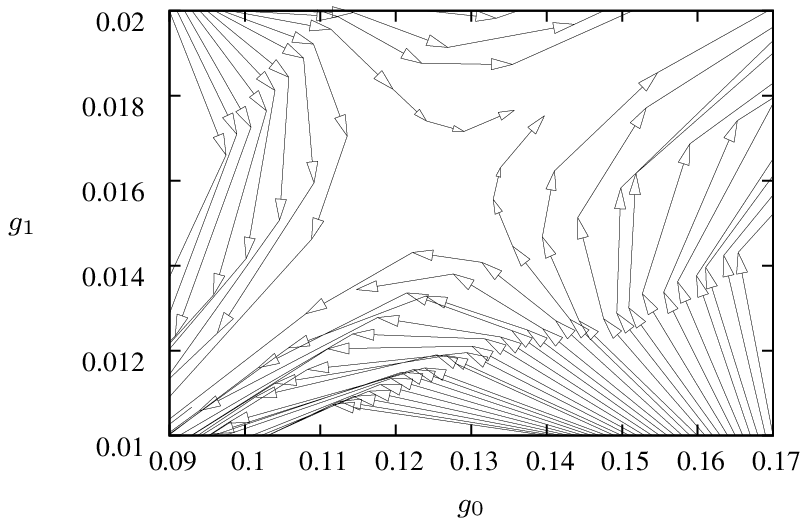}}\\
\scalebox{0.98}{\includeEPSTEX{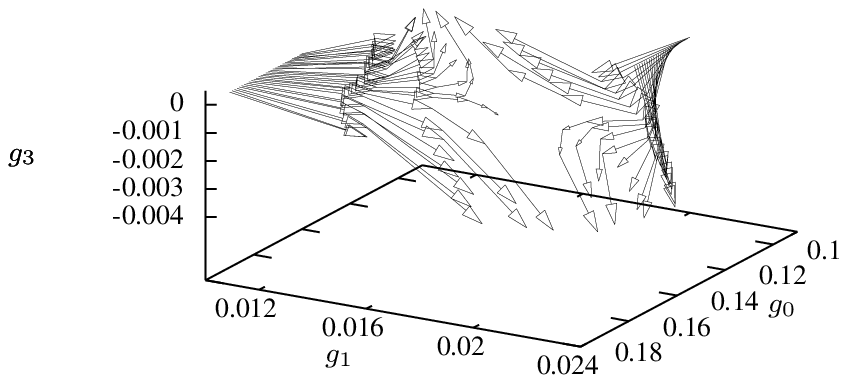}}
\scalebox{0.98}{\includeEPSTEX{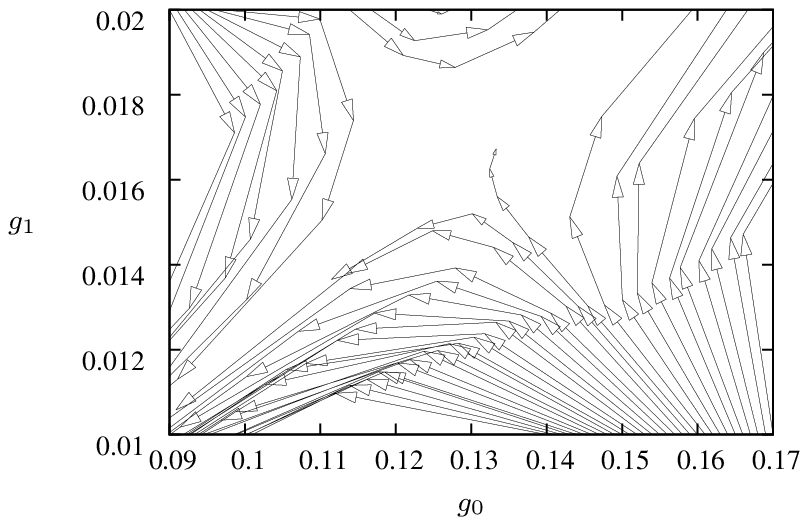}}\\
\scalebox{0.98}{\includeEPSTEX{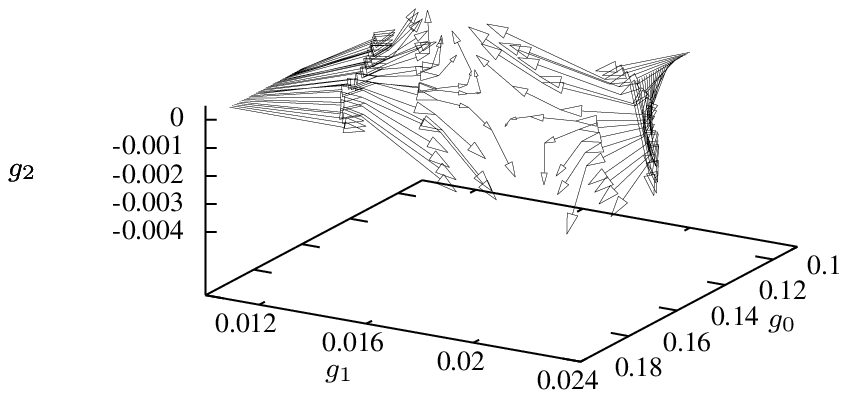}}
\scalebox{0.98}{\includeEPSTEX{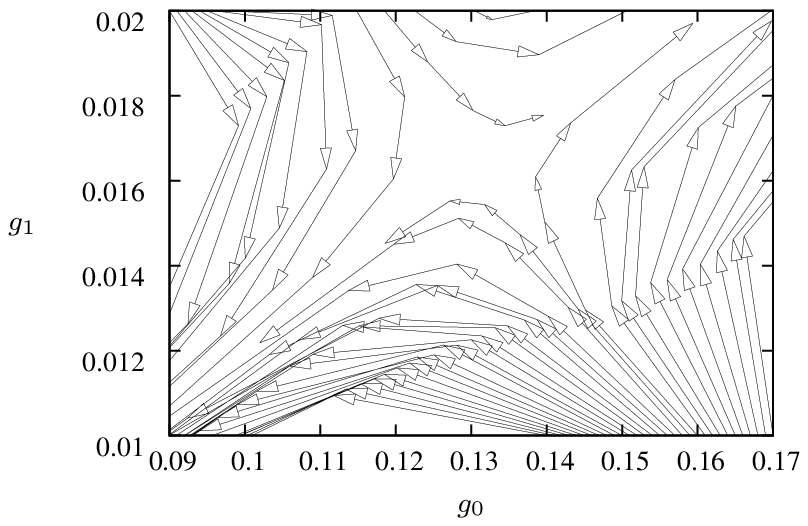}}
\caption{For higher order truncations with operators $\{S_0, S_1, S_2\}$ in 
the upper panel, $\{S_0, S_1, S_3\}$ in the center panel and 
$\{S_0, S_1, S_2, S_3\}$ in the lower panel the fixed point structure 
of the resulting flow diagram remains the same as for the $1 \rightarrow 1$ 
and $2\rightarrow2$ truncation.}
\label{fig:flowhigher}
\end{figure*}
The blockspin transformation is optimized in the same way as for the action
with a single parameter. Our choice for the parameters is $c_0=3.1$ and 
$c_1=2.5$ and it leads to a correlation length ratio of around $2$ in 
the vicinity of the fixed point. Note that this choice for the parameters is not unique if we only tune the correlation length to the desired value. In general we have to
consider higher correlation functions as well. Below we will also discuss other choices for the parameters and its 
influence on quantitative features of the flow diagram as for example the position of the fixed point or critical exponents.
Nevertheless as in the one parameter case the choice of the parameters does not change the qualitative flow diagram.
Again, we detect a high temperature fixed point (HT FP) at zero coupling in the lower left corner 
as well as a low temperature fixed point (LT FP) at infinite 
coupling. Also a non-trivial fixed point (NG FP) in the center 
of the flow diagram is clearly visible. The values of the couplings at the fixed point, 
$g_0=0.119(1)$, $g_1=0.0164(2)$, can be determined from Fig.~\ref{fig:flowop22} 
(lower panel).  As expected, the `velocity' along a trajectory gets small in the 
fixed point regime. Furthermore, we find that the position of the fixed 
point in this two parameter truncation is almost independent of the 
lattice volume. But, in contrast to the one-parameter truncation, 
it depends strongly on the constant $c_0$ and to a lesser degree on the
remaining constants. A change of $c_0$ results in
a displacement of the fixed point along the critical line.\\
The flow diagram is split by a separatrix which defines the
critical line (CL) extending from the lower right to the 
upper left corner. Trajectories that lie above this line will flow into 
the low temperature fixed point while trajectories below this line flow 
into the high temperature fixed point. This indicates a relevant 
direction analogous to the simple
one-parameter truncation of the preceding section. The second direction
though is an irrelevant one and the corresponding eigenvalue of 
the stability matrix is negative. The single 
trajectory that is identical with the critical line 
will flow into the non-trivial fixed point, either from below or above.
The critical line is the intersection of the critical hypersurface in
general theory space with the $g_0$-$g_1$ plane that constitutes
our truncation. From the traditional lattice perspective, the critical 
line corresponds 
to a fine-tuned set of bare couplings $(g_0$, $g_1)$ at different 
UV cutoffs. Starting a simulation on the critical line results in a 
measurement of the critical physics at the non-trivial fixed point and 
is generically used to take the continuum limit, since the lattice spacing in units 
of the correlation length becomes small as the critical point is 
approached.\\
There exists another interesting line which connects all three fixed points
and acts as an attractor for the RG trajectories. It is called
the renormalized trajectory (RT) and singles out a unique
trajectory that defines a theory that is both IR and UV complete, 
starting at the non-trivial fixed point in the UV and flowing into the
high temperature or low temperature fixed point in the IR. As expected, the
RT does not attract the trajectories in the vicinity of the high temperature 
fixed point, where the fixed point behaviour dominates
\footnote{For this reason the matching method is not applicable in the vicinity of the 
high temperature fixed point since it relies on the assumption that the 
trajectories approach the renormalized trajectory within a few 
RG steps \cite{Hirsch1983}.}.\\
Starting on the $g_0$ axis, which corresponds to the usual lattice
action of the Heisenberg ferromagnet, and integrating out all 
fluctuations, one can only reach either one of the trivial fixed points 
or the non-trivial fixed point. In this sense, it is
legitimate to consider them as \emph{infrared fixed points}. From
universality arguments one expects that the non-Gau{\ss}ian fixed point corresponds 
to the well-known Wilson-Fisher fixed point of the linear sigma model.
We find that a similar structure to our results emerges in this model \cite{Bohr2000}.\\
But the Heisenberg ferromagnet is an effective theory that is 
well defined only for a finite UV cutoff, in contrast to
\emph{asymptotically safe theories} that are defined on all scales.
Fundamental field theories correspond to theories on the renormalized 
trajectory and the direction of the renormalization group flow shows 
that the non-trivial fixed point governs the ultraviolet 
physics of these theories. Thus, this non-trivial fixed point acts
as an \emph{ultraviolet fixed point} of the RG flow.\\
For the asymptotic safety scenario to hold, the number
of relevant directions at the non-Gau{\ss}ian fixed point must be 
finite. Hence we proceed to determine the flow diagram for the $3\rightarrow 3$ 
and $4\rightarrow 4$ truncation, which include the operators 
$\{S_0, S_1, S_2\}$, $\{S_0, S_1, S_3\}$ and $\{S_0, S_1, S_2, S_3\}$
respectively. An overview over the full flow diagram for 
the operators $\{S_0, S_1, S_2\}$ is presented in Figure \ref{fig:flowop33} and
it is evident that only irrelevant directions are added to the 
truncation. The global structure of the flow diagram is similar to
the $2\rightarrow 2$ truncation and shows two trivial IR fixed points
and one non-trivial UV fixed point. 
Figure \ref{fig:flowhigher} (upper panel) shows a detailed view of
the the fixed point regime. The fixed point is located at 
$(g_0,g_1,g_2)=(0.13(1),0.016(1),-0.0015(5))$.
In the center panel of Figure \ref{fig:flowhigher} the $3\rightarrow3$ 
truncation with operators $\{S_0, S_1, S_3\}$ is presented.
The resulting flow diagram is again very similar and we find that even
the position of the fixed point at 
$(g_0,g_1,g_3)=(0.13(1),0.016(1),-0.0015(5))$ matches the prior result
within the resolution of the flow diagram.
Finally Fig.~\ref{fig:flowhigher} 
(lower panel) shows the results for the $4\rightarrow4$ truncation.
Again the fixed point structure remains unchanged. In this truncation 
the position of the fixed point is at $(g_0,g_1,g_2,g_3)=(0.13(1),0.016(1),-0.0015(5),-0.0015(5))$.
In conclusion, we observed that the fixed point structure does not change 
if we add further operators. We always find just one relevant direction
at the non-Gau{\ss}ian fixed point. In addition the position of the fixed point 
is stable against including the higher derivative operators $S_2$ and $S_3$.
This clearly points to the existence of a non-Gau{\ss}ian
fixed point of the RG transformation and thus we are led to believe
that the asymptotic safety scenario applies to the \OD nonlinear sigma 
model in three dimensions.

\subsection{Critical exponents}
\label{critExp}

\begin{figure}[htb]
\scalebox{1}{\includeEPSTEX{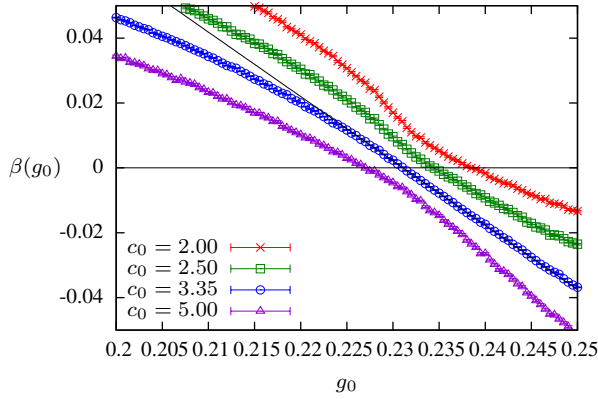}}
\caption{$\beta$ function in the vicinity of the fixed point in the 
$1\rightarrow 1$ truncation for the \OD-model and various values of $c_0$.}
\label{fig:beta_op11_3}
\end{figure}

\begin{figure}[htb]
\scalebox{1}{\includeEPSTEX{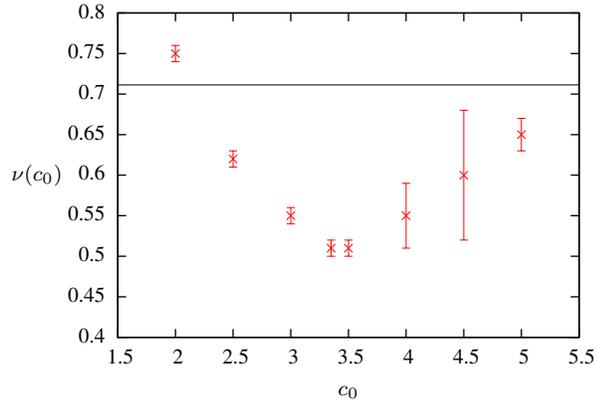}}
\caption{The critical exponent $\nu$ is shown as a function of $c_0$.}
\label{fig:critExp_c0}
\end{figure}

\begin{figure}[htb]
\scalebox{1}{\includeEPSTEX{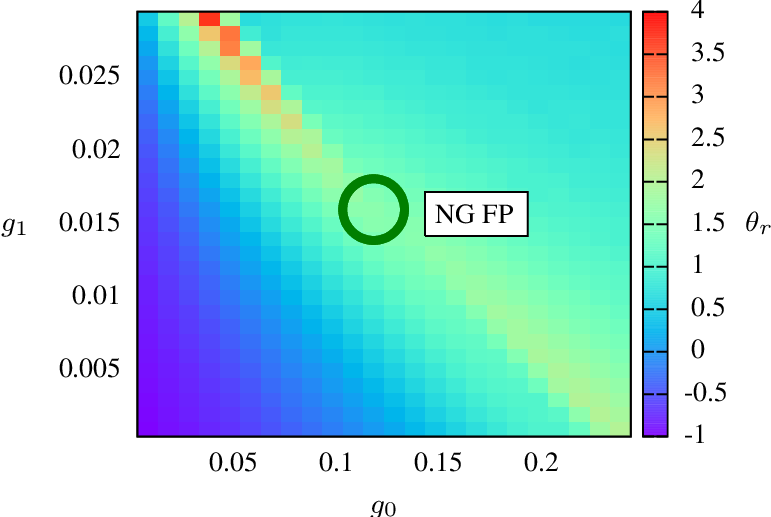}}\\
\scalebox{1}{\includeEPSTEX{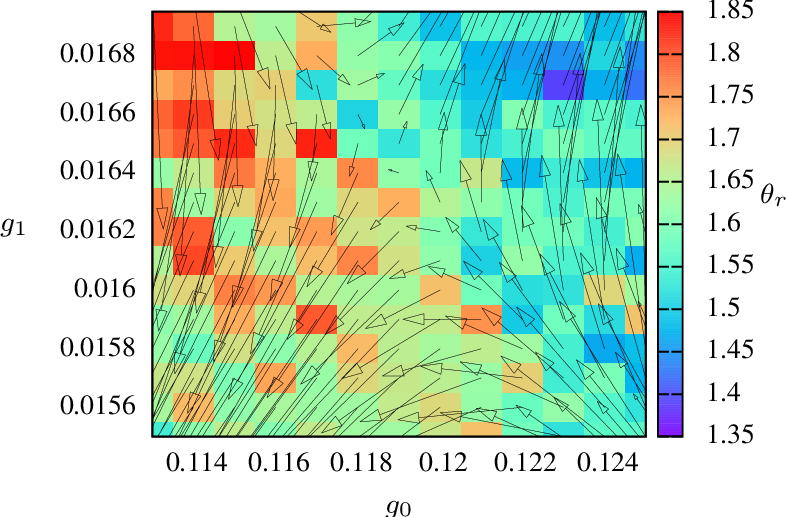}}
\caption{The critical exponent $\nu$ is related to the eigenvalue 
$\theta_r$ of the stability matrix corresponding to the relevant 
direction. The RG parameters for this flow diagram are $c_0=3.1$ and $c_1=2.5$.}
\label{fig:critexp_r}
\end{figure}

\begin{figure}[htb]
\scalebox{1}{\includeEPSTEX{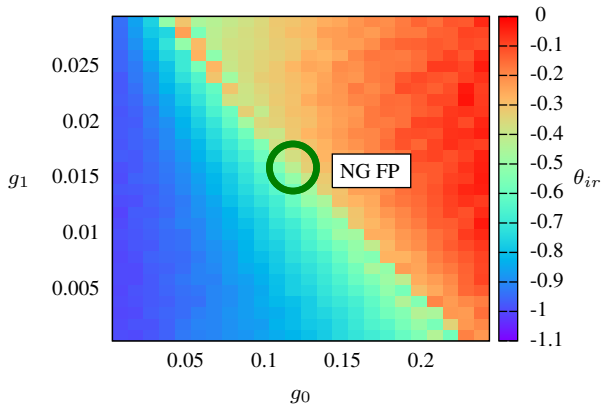}}\\
\caption{The critical exponent corresponding to the irrelevant direction is negative at the high temperature and non-Gau{\ss}ian fixed point. The RG parameters for this flow diagram are $c_0=3.1$ and $c_1=2.5$.}
\label{fig:critexp_ir}
\end{figure}

\noindent Following the universality hypothesis, it is generally 
assumed that the linear and nonlinear \ON models are in the same
universality class, since they have the same range of interaction 
and symmetries. This assumption is supported by several computations 
based on very different approximations, cf. for 
instance \cite{Zinn,Ballesteros,Butera1997} or the overviews
\cite{Pelissetto,Wipf:2013vp}.\\
Furthermore, critical exponents are universal, 
in contrast to the position of the fixed point, and this allows us to compare
our results to the functional RG studies of the nonlinear
\ON models in \cite{Flore2012}.
Here we restrict ourselves to the scaling properties of the 
correlation length, described by the exponent $\nu$, since it is 
directly related to the relevant eigenvalue $-\theta_r$ 
of the stability matrix by $\nu = \theta_r^{-1}$.\\
Using the simple $1\rightarrow 1$ truncation, the inverse of the 
thermodynamic critical exponent $\nu$ corresponds to the negative slope of the lattice beta 
function in the vicinity of a fixed point, depicted in Fig.~\ref{fig:beta11}. 
As expected, we find the trivial values  $\nu\approx-1$  and $\nu\approx1$ for the 
high-temperature and low-temperature fixed points, respectively. 
These values are almost independent of $c_0$. For the non-trivial fixed point, 
on the other hand, the value depends on the choice for $c_0$, and
this is shown in Fig.~\ref{fig:beta_op11_3}.
For the optimal constant $c_0=3.35$ we read off the critical exponent 
$\nu(1\rightarrow 1)=0.51(1)$ 
for $N=3$, which is to be compared with the value $0.7112(5)$ 
in \cite{Campostrini2002} obtained from 
a dedicated Monte Carlo simulation combined with high-temperature expansions.
In Fig.~\ref{fig:critExp_c0} the critical exponent is shown 
as a function of $c_0$. 
Again one sees that a careful optimization of the blockspin transformation 
is important in order to extract accurate results for the critical exponents.\\
The next improvement is to allow 2 operators in the effective action, 
denoted as $1\rightarrow 2$ truncation. The critical exponent
is determined as the negative slope of the projected $\beta$-function on the 
$g_0$ axis at the position of the fixed point for the $1 \rightarrow 1$ 
truncation. For the optimized $c_0=c_0^{\text{opt}}$ we obtain 
$\nu(1\rightarrow 2)=0.55(2)$. This is already significantly closer to 
the expected value compared with the simple $1\rightarrow 1$ 
truncation.\\
We can further improve our estimate by moving on to the $2\rightarrow 2$ truncation.
Depicted in Figure \ref{fig:critexp_r} (upper panel) is the eigenvalue $\theta_r$ 
of the matrix (\ref{eq:stabmatw}), which at a critical point becomes 
the stability matrix, and 
again it takes the trivial values at the high temperature or low temperature fixed 
point. 
While the plot shows strong variations of the eigenvalue at 
the upper left and lower right corner of the parameter space, it 
becomes smooth in the vicinity of the non-trivial fixed point, 
see Fig.~\ref{fig:critexp_r} (lower panel). 
From an average over the fixed point region we
obtain the value of $\nu(2\rightarrow 2)=0.62(3)$, which already 
deviates less than $15\%$ from the literature value. We stress 
that in the present work we are mainly concerned with the flow diagram 
and fixed point structure of non-linear \ON-models such that our method
is not to be seen as a replacement of dedicated high-precision 
Monte Carlo determination of critical exponents. 
It is however possible to estimate these quantities in addition to the 
flow diagram with a reasonable precision.\\
In the $2\rightarrow 2$ truncation we can also extract the critical exponent corresponding
to the irrelevant direction of the flow, see Fig.~\ref{fig:critexp_ir}.
It takes the trivial value $\theta_{ir}=-1$ at the high temperature fixed point
and $\theta_{ir}\approx -0.44$ at the non-Gau{\ss}ian fixed point.
In order to check for the stability of our method we calculated the critical exponent also for the RG parameters $c_0=3.4$ and $c_1=1.0$. We obtained the value
$\nu=0.65(3)$. Within statistical errors this agrees with the value for $\nu$ obtained before.\\
For the $3\rightarrow 3$ truncation we set the RG parameter belonging to the additional operator to zero, i.e. $c_0=3.1$, $c_1=2.5$ and $c_2=0$.
In this truncation we obtain three critical exponents:
\begin{equation}
\begin{aligned}
\theta_r=& 1.57(5)\,,\\ \theta_{ir}^1=& -0.52 + 0.05\ii\,,\\ \theta_{ir}^2=&-0.86 - 0.05\ii.
\end{aligned}
\end{equation}
The exponent of the correlation length is then $\nu(3\rightarrow 3)=0.64(3)$. Within 
statistical errors  this is almost no improvement compared to the $2\rightarrow 2$ 
truncation.\\
Our analysis of the critical exponents indicates that the high temperature fixed point has 
only irrelevant directions, 
i.e. all critical exponents are negative. The exponents corresponding to the operators $S_0$ and $S_1$ take the value $\theta_{0,1}=-1$.
The non-Gau{\ss}ian UV fixed point has one positive critical exponent, while the other critical exponents are negative.
This again verifies the asymptotic safety scenario for the nonlinear sigma model in three dimensions.
Table \ref{tab:critSummary} summarizes our results for the critical exponents.
\begin{table}[htb]
\begin{tabular}{|c|c|c|}\hline
Method & $\nu$ & $\nu/\nu_{MC}$\\ \hline
$1\rightarrow1$ trunc. ($c_0=3.35$) & $0.51(1)$ & $\sim 0.71$\\ \hline
$1\rightarrow2$ trunc. ($c_0=3.35$) & $0.55(2)$ & $\sim 0.77$ \\ \hline
$2\rightarrow2$ trunc. ($c_0=3.1,\, c_1=2.5$)& $0.62(3)$ & $\sim 0.87$ \\ \hline
$2\rightarrow2$ trunc. ($c_0=3.4,\, c_1=1.0$)& $0.66(4)$ & $\sim 0.93$ \\ \hline
$3\rightarrow3$ trunc. ($c_0=3.1,\, c_1=2.5,\, c_2=0$)& $0.64(3)$ & $\sim 0.90$ \\ \hline
FRG \cite{Flore2012} & $0.704$ & $\sim 0.99$\\ \hline
MC \cite{Campostrini2002} & $0.7112(5)$ & $1$ \\ \hline
RG \cite{Antonenko1998} & $0.706$ & $\sim 0.99$ \\ \hline
HT \cite{Butera1997} & $0.715(3)$ & $\sim 1$\\ \hline
\end{tabular}
\caption{
Results for the critical exponent $\nu$ for different truncations and $N=3$ compared to the very precise results of the Monte Carlo estimate MC.
}
\label{tab:critSummary}
\end{table}
For comparison we also show results obtained with Monte Carlo 
simulations (MC), high temperature expansion (HT), RG expansion (RG) 
and functional RG (FRG). With increasing truncation order our results 
approach the very precise values obtained with other methods, indicating 
that our derivative expansion converges to the correct results. For 
even higher truncations the computation of critical exponents becomes 
very time consuming and the statistical errors 
become larger than the deviation from the values in the literature.
Furthermore the optimization of the blockspin transformation becomes
increasingly difficult. Nevertheless results are good enough to
show that the non-Gau{\ss}ian UV fixed point indeed belongs to a well-known
class of second order phase transitions.

\section{The large $N$ limit}\label{largeNResults}

\begin{figure}[htb]
\scalebox{1}{\includeEPSTEX{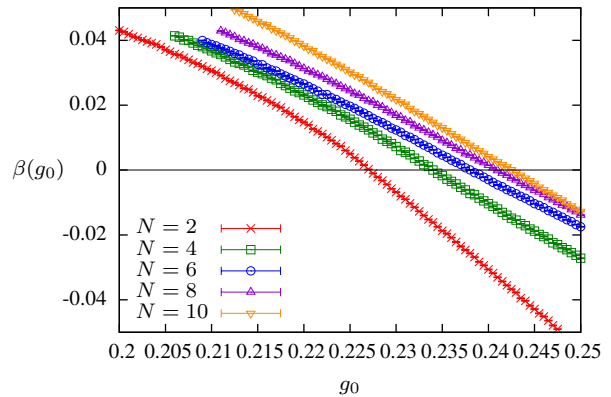}}
\caption{Shown is the $\beta$ function in the one parameter truncation for various $N$ and optimized blockspin transformation.}
\label{fig:beta_op11_N}
\end{figure}

\begin{figure}[htb]
\includeEPSTEX{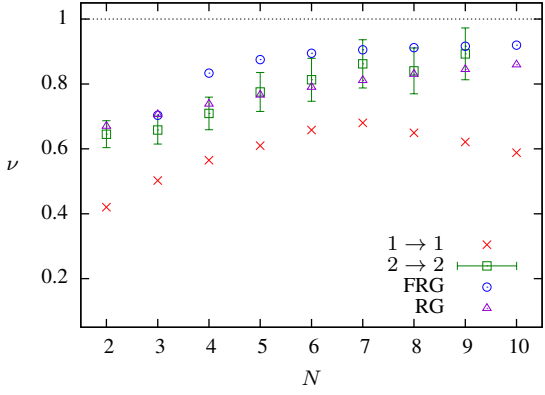}
\caption{The critical exponent $\nu$ is shown for the $1\rightarrow 1$ 
and $2\rightarrow 2$ truncation depending on $N$. 
We compare our data to results using the Functional 
RG \cite{Flore2012} and RG expansion \cite{Antonenko1998}. In the large 
$N$ limit we expect $\nu=1$.}
\label{fig:nulargeN}
\end{figure}

\begin{figure}[htb]
\scalebox{0.95}{\includeEPSTEX{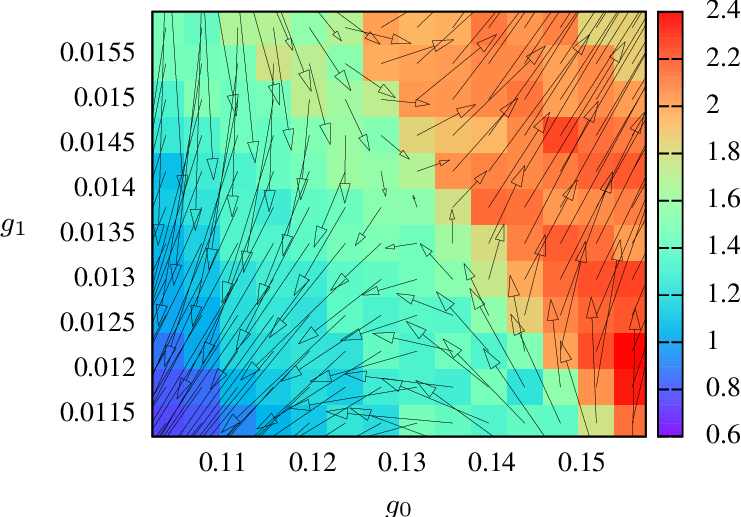}}\\
\scalebox{0.95}{\includeEPSTEX{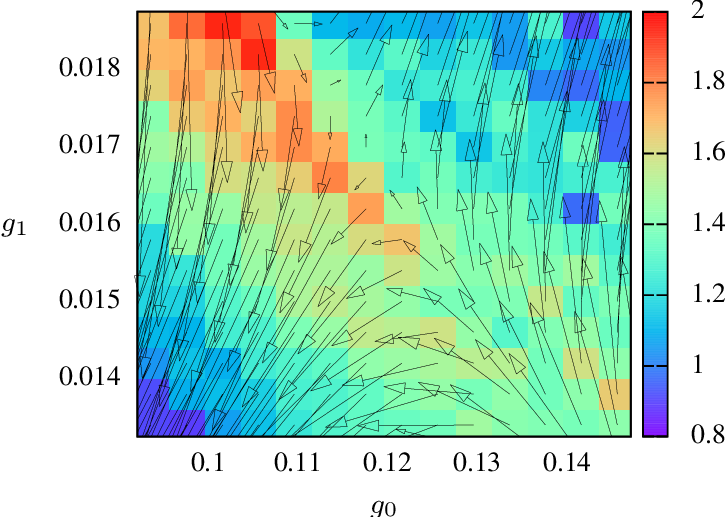}}\\
\scalebox{0.95}{\includeEPSTEX{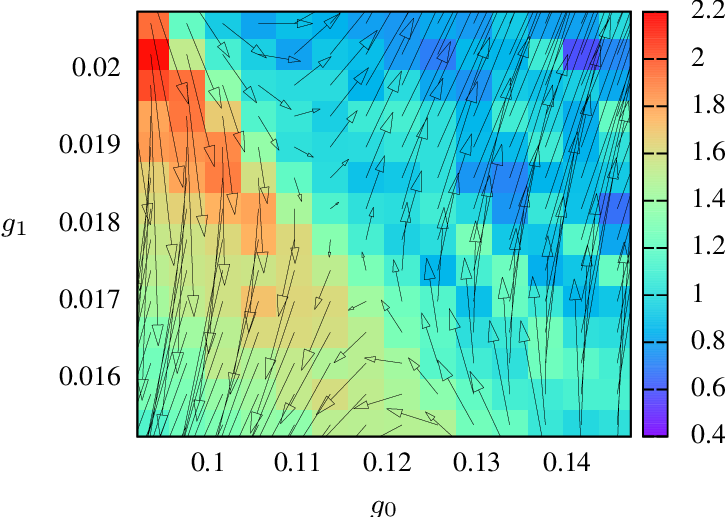}}
\caption{Flow diagram using the $2\rightarrow 2$ truncation in three
dimensions for O(2), O(4) and O(6)-models. The global structure is
the same as for the \OD-model.
The background color encodes the eigenvalue
$\theta_r$ of the matrix in (\ref{eq:stabmatw}), which near
a critical point is related to the critical exponent of the 
correlation length.}
\label{fig:flowop22_24}
\end{figure}

\noindent For large values of $N$ we can compare our 
results with those from the analytical large $N$ and RG expansions in  
\cite{Okabe1978} and \cite{Antonenko1998}, respectively.
In Fig.~\ref{fig:beta_op11_N} the $\beta$-function in the $1\rightarrow1$ truncation 
is shown for different $N$ at the optimized value for $c_0(N)$.
For every value of $N$ a non-trivial fixed point exists, but the slope at the 
fixed point changes. In order to connect to the large $N$ limit, we repeat the computation
of the critical exponent $\nu$ in the simple $1\rightarrow 1$ truncation
for $N$ up to $10$. The results are shown in Fig.~\ref{fig:nulargeN}.
Starting from $N=2$, 
where the estimate deviates from the comparative RG
data by $\nu/\nu_{RG}\approx 40\%$, we see a significant improvement for 
intermediate $N<8$. However, going to even larger $N$, the behaviour 
changes and our results significantly underestimate the correct values. 
It is evident that we do not reproduce the analytically 
known result of $\nu=1$ for $N\rightarrow\infty$.
\begin{figure}[htb]
\includeEPSTEX{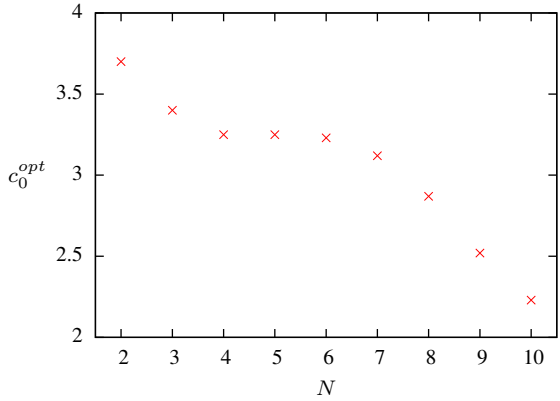}
\caption{The optimization constant $c_0$ in the $1\rightarrow 1$
truncation is shown for various $N$.}
\label{fig:clargeN}
\end{figure}
This change of behaviour is not only visible in the critical exponents but 
also shows up in the value of the optimization constant $c_0$.
From the perturbative analysis \cite{Hasenfratz1984}, we know that for 
large $N$ the RG parameter is proportional to $\frac{N}{N-1}$, i.e. we 
expect $c_0(N)$ to become constant for large $N$.
Indeed in Figure \ref{fig:clargeN} we see a plateau for intermediate values of $N$.
Unfortunately, for $N\gtrsim 7$ the optimization
constant decreases rapidly. We interpret this unexpected behaviour as a
breakdown of our simple one parameter truncation for large $N$. If the 
effective action does not capture the relevant physics anymore, then we 
should not expect to find reliable values for the critical exponents.
Although we can tune the ratio of two-point functions to the desired 
value, higher correlation functions should indicate that, within our 
truncation, the IR physics changes under the RG transformation.\\
We might try to improve the situation by including higher order
operators. For the $2\rightarrow 2$ truncation we calculated the critical 
exponents up to $N=9$ and actually see a 
significant improvement over the $1\rightarrow1$ truncation, 
see Fig.~\ref{fig:nulargeN}. It turns out that, compared to the literature, we get the best results if we set the
RG parameter $c_0$ to the values obtained in the simplest truncation for $N\leq 6$ and to the plateau value for $N>6$.
For the second operator we choose $c_1=1.0$. We checked that the ratio of the correlation length is approximately $2$ in the vicinity of
the fixed point for this set of parameters. 
In Figure \ref{fig:flowop22_24}, we show
that for different $N$ the general structure of the flow diagram 
persists. Only the non-universal location of the non-Gau{\ss}ian fixed point 
varies.
\\
Unfortunately the fine-tuning of the RG parameter and the computation 
of critical exponents becomes increasingly difficult for even 
larger $N$. We again observe that for $N>9$ our truncation breaks down 
and additional operators are needed to obtain reliable results for the 
critical exponents. Nevertheless the fixed point structure itself 
remains stable.
Our final results are compiled in Table 
\ref{tab:nulargeN} and Figure \ref{fig:nulargeN}.
\begin{table*}[htb]
\begin{tabular}{|c|c|c|c|c|c|c|c|c|c|} \hline
N & $2$ & $3$ & $4$ & $5$ & $6$ & $7$ & $8$ & $9$ & $10$  \\ \hline \hline
$1\rightarrow 1$ truncation      & $0.42$ & $0.51$ & $0.57$ & $0.63$ & $0.65$ & $0.68$ & $0.65$ & $0.62$ & $0.58$ \\ \hline
$2\rightarrow 2$ truncation & $0.64(4)$ & $0.66(4)$ & $0.71(5)$ & $0.78(6)$ & $0.81(6)$ & $0.86(7)$ & $0.84(7)$ & $0.89(8)$ & - \\ \hline
FRG \cite{Flore2012}          & -       & $0.704$ & $0.833$ & -       & $0.895$ & - & $0.912$ & - & $0.920$  \\ \hline
HT exp. \cite{Butera1997} & $0.677(3)$ & $0.715(3)$ & $0.750(3)$ & - & $0.804(3)$ & - & $0.840(3)$ & - & $0.867(4)$  \\ \hline
RG exp. \cite{Antonenko1998} & $0.607$ & $0.706$ & $0.738$ & $0.766$ & $0.790$ & $0.811$ & $0.830$ & $0.845$ & $0.859$  \\ \hline
\end{tabular}
\caption{
Results for the critical exponent $\nu$ for different N obtained using different methods.
}
\label{tab:nulargeN}
\end{table*}

\section{Conclusions}
\label{Con}

\noindent
We have discussed and applied a method that allows to compute the global flow 
diagram of a model from numerical simulations. In contrast to the
MCRG matching technique, our method does not need exponentially large 
lattices and works even in the vicinity of a Gau{\ss}ian fixed point,
where the renormalized trajectory no longer acts as an attractor for
the RG flow. Furthermore, we have shown that systematic uncertainties
from a truncation of the effective action can be mitigated efficiently
by an optimization of the RG transformation.\\
The nonlinear sigma model is asymptotically free in two dimensions and
we have reproduced the expected structure of the flow diagram, 
showing two trivial fixed
points corresponding to the behaviour at very low and 
very high temperature, already using the simplest
possible truncation that only includes a nearest-neighbor interaction.
Using a two-operator truncation, we have clarified the role of the
finite volume behaviour on the flow diagram and argued 
that an additional non-trivial fixed point is a lattice 
artifact.\\
It has long been known that the three-dimensional \OD-model shows a 
second-order phase transition that separates a phase of broken \ON 
symmetry and a symmetric phase. We have shown that this phase transition
corresponds to an ultraviolet fixed point with \emph{one} relevant direction
by using a truncation that includes all (allowed by symmetry) operators 
up to fourth order in the momentum.
It is possible to define a theory along the renormalized trajectory
that is IR- and UV-complete. We conclude that the asymptotic safety
scenario is fulfilled and the model is renormalizable in a 
non-perturbative setting.\\
While the general structure of the flow diagram does not depend on the 
specific RG scheme the critical exponents vary since the systematic 
error depends on the specific optimization constant. We find that our 
method is able to predict the critical exponents within a reasonable
accuracy but can not compete to designated high precision 
MC-techniques that are free of truncation errors 
\cite{Campostrini2002}. We find that our estimates for the critical
exponents improve for larger truncations but fail to reproduce the
exact $N\rightarrow\infty$ limit.\\
Using functional renormalization group techniques, the
full flow diagram for the present model was obtained already in an earlier 
publication \cite{Flore2012}. We find that the qualitative structure of 
the flow diagrams are the same. However, the MCRG method
is more stable than the FRG method and leads to more robust 
results for different truncations. In particular, we do not find a sudden 
disappearance of the non-trivial fixed point for a certain truncation
including the operator $S_2$ (\ref{eq:ops}). 
Furthermore, we stress that lattice techniques provide
the opportunity to obtain additional information beyond the chosen 
truncation by a direct measurement of the Green's functions. 
We have used this knowledge to determine the 
optimal constants in the improved RG transformation. In addition 
we compared the location of the critical point, determined by the
susceptibility of the order parameter, to the location of the fixed
point, determined by the zero crossing of the beta function
and hence amendable to truncation errors. We find that these points do
not coincide in general. For the simplest truncation we observe a small 
deviation even for the optimal value of the RG constant.
For higher truncations, the fixed point location matches the critical 
surface within statistical errors. Another interesting observation is 
that the $\beta$ function in the lowest truncation for two and three 
dimensions does not depend on the lattice size.\\
Our method can be generalized to other systems, especially including 
fermionic degrees of freedom, and thus allows to determine the more 
complex flow diagrams of e.g. the Thirring model \cite{Gies2010}.
The method might also be used to study lattice quantum gravity
\cite{Ambjorn:2001cv,Ambjorn:2013tki} where it is difficult to define 
observables that capture the infrared physics of the theory. 
In contrast to the matching technique, the method used in the present
work does not rely on the computation of correlation functions but only 
on an appropriate RG transformation that acts directly on the spacetime 
triangulations used. 

\begin{acknowledgments}
\noindent
This work was supported by the DFG-Research Training Group ''Quantum- and 
Gravitational Fields'' GRK 1523 and DFG grant Wi777/11 and by the Helmholtz 
International Center for FAIR within the LOEWE initiative of the State of Hesse.
We thank Raphael Flore for his active collaboration and Jens Braun, Axel Maas, 
Roberto Percacci, Martin Reuter, Omar Zanusso and 
Luca Zambelli for interesting discussions or useful comments. 
Simulations were performed on the Omega HPC cluster at the University of Jena
and the LOEWE-CSC at the University of Frankfurt.
\end{acknowledgments}

\appendix 

\section{Two Alternative Formulations of a Fourth-Order Derivative Expansion}
\label{expRel}
In this article we study the full fourth order derivative expansion of 
the theory, formulated in terms of explicitly constrained variables 
$\bphi \in \mathbb{R}^N$ with $\bphi\cdot\bphi=1$. 
In order to compare the results with previous studies of the same system by means of the Functional Renormalization Group (FRG) \cite{Flore2012}, 
one has to know the relation between both parametrizations of the action functional. 
The FRG computations in \cite{Flore2012} were performed for the covariant formulation 
\begin{equation}
\begin{aligned}
\label{covariantAction}
 \Gamma[\varphi] = &  \frac{1}{2} \int d^3 x~\zeta h_{ab} \pmu \varphi^a \pMu \varphi^b \\
  &+ \alpha h_{ab} (\Pmu\pMu \varphi)^a (\Pnu\pnu \varphi)^b \\
  &+ L_1  (h_{ab} \pmu \varphi^a \pnu \varphi^b)^2 \\
  &+ L_2  (h_{ab} \pmu \varphi^a \pMu \varphi^b)^2\,,
\end{aligned}
\end{equation}
in terms of unconstrained fields $\varphi\in \mathbb{R}^{N-1}$, where $(\Pmu\pMu \varphi)^a = \partial^2 \varphi^a + \Gamma^a_{~bc}\pmu\varphi^b\pMu\varphi^c$ and $\Gamma^a_{~bc}$ is the Christoffel symbol corresponding to the metric $h_{ab}(\varphi)$.
In order to determine the relation between \eqref{covariantAction} and \eqref{contAction}, one can choose stereographic coordinates,
\begin{equation}
 h_{ab} = \frac{\delta_{ab}}{(1+\varphi^2)^2}\quad\text{with}~\varphi^2 = \sum_{a=1}^{N-1} \varphi^a \varphi^a\,,
\end{equation}
for an unconstrained parametrization of \eqref{covariantAction} and apply an inverse 
stereographic projection,
\begin{equation}
 \varphi^a = \frac{\phi^a}{1+\phi^N}\quad \text{for}~i=1,..,N-1\,,
\end{equation}
such that 
\begin{equation}
\begin{aligned}
 \Gamma[\varphi(\phi)] =& \frac{1}{2}\,\frac{\zeta}{4}\,\pmu \phi \pMu \phi + \frac{1}{2} \,\frac{\alpha}{4}\, \partial^2\phi \partial^2\phi \\
 &+ \frac{1}{2}\,\frac{L_1}{16}\,(\pmu\phi\pnu\phi)^2 \\
 &+ \frac{1}{2}\,\frac{L_2\!-\! 4\alpha}{16}\,(\pmu\phi\pMu\phi)^2\,.
\end{aligned}
\end{equation}
A comparison with \eqref{contAction} yields
\begin{align}
 g_0 = \frac{\zeta}{4}\,,\quad g_1 = \frac{\alpha}{4}\,,\quad g_2 = \frac{L_1}{16}\,,\quad g_3 = \frac{L_2-4\alpha}{16}\,.
\end{align}

\section{The LHMC algorithm}

\label{sec:algo}
\noindent
In the case of nonlinear sigma models with only the standard interaction 
term $S_0$, cluster algorithms have proven to be the most efficient way 
to update the scalar field in Monte-Carlo simulations. In its original 
version, the cluster algorithm assumes that only nearest neighbor 
interactions are present and hence is not directly applicable
in the presence of higher derivative operators. Thus
we employ a local version of the hybrid Monte-Carlo algorithm (LHCM)
where single site variables are evolved in an HMC algorithm. This ansatz relies on
local interactions and is applicable theories without dynamical fermions.
The formulation is given entirely in terms of \SON-Lie-group and 
Lie-algebra elements, see also \cite{Wellegehausen:2011sc}. To update the normalized scalar field we set
\begin{equation}
 \Phi_x=\gO_x\Phi_0\quad\hbox{with}\quad \gO_x\in \text{\SON}
\end{equation}
and constant $\Phi_0$. The change of variables $\Phi_x\to \gO_x$
converts the  induced measure on $S^{N-1}\subset R^N$ into the Haar measure of
\SON. Without interaction the rotation matrices $\gO_x$ will evolve freely on
the group manifold \SON. The \emph{free evolution} on a semisimple group is
the Riemannian geodesic motion with respect to the Cartan-Killing metric
\begin{equation}
ds^2 \propto \tr\left(d\gO \gO^{-1}\otimes d\gO\gO^{-1}\right).
\end{equation}
The LHMC dynamics may be naturally derived from a Lagrangian of the 
form
\begin{equation}
L=-\frac{1}{2}\sum\limits_x  \tr\left(\dot\gO_x\gO^{-1}_x\right)^2-S[\,\gO],
\end{equation}
where `dot' denotes the derivative with respect to the fictitious time 
parameter $\tau$. The Lie-algebra valued pseudo-momenta conjugated to 
the site variable $\gO_x$ are given by
\begin{equation}
\begin{aligned}
\algebra{P}_x&= \frac{\partial L}{\partial
\big(\dot{\gO}_{x}\gO_{x}^{-1}\big)}=-\dot\gO_x\gO_x^{-1}.
\label{hmcequations1}
\end{aligned}
\end{equation}
The Legendre transform yields the following pseudo-Hamiltonian
\begin{equation}
H=-\frac{1}{2}\sum_x \tr \algebra{P}_x^2
+S[\,\gO].
\end{equation}
Note that for $\gO_x\in$ \SON the momenta are antisymmetric
such that the kinetic term is positive. The equations of motion for the
momenta are obtained by varying the Hamiltonian,
\begin{equation}
\label{HMChamiltonian}
\begin{aligned}
\delta H &=-\sum_x \tr \algebra{P}_x \big\{\dot{\algebra{P}}_x- F_x\big\} \quad
\text{and} \quad F_x=\delta S[\,\gO].
\end{aligned}
\end{equation}
In the simplest case of only nearest neighbor interactions the \emph{force} is
given by
\begin{equation}
\label{HMCForce}
\begin{aligned}
F_x&=g_0\,\Phi_x\Big(\sum\limits_{x,\mu} \,\Phi_{x+\mu}\Big)^\trnsp\,.
\end{aligned}
\end{equation}
The variational principle implies that the projection of the terms between
curly brackets onto the Lie-algebra $\algebra{so}(N)$ vanishes,
\begin{equation}
\dot{\algebra{P}}_x=F_x\big\vert_{\algebra{so}(N)}.\label{hmcequations2}
\end{equation}
There is a freedom of choice of $F$ and we determine it by a projection on a trace-orthonormal basis $\{T_a\}$ of $\algebra{so}(N)$.
Then the LHMC equations read
\begin{equation}
 \dot{\gO}_x=-
\algebra{P}_{x}\gO_x\quad \text{and} \quad
\dot{\algebra{P}}_{x}=\sum \limits_b \tr \left(F_{x}T_b\right)T_b\,.
\end{equation}
To solve these equations of motion numerically, we employ a time reversible leap
frog integrator which uses the integration scheme 
\begin{equation}
 \begin{aligned}
\algebra{P}_{x}(\tau+\ft12\delta\tau)&=\algebra{P}_{x}(\tau)
+\ft12\delta\tau\, \dot{\algebra{P}}_{x}(\tau)\\
\gO_{x}(\tau+\delta\tau)&=\exp\left\{-\delta\tau\,
\algebra{P}_{x}(\tau+\ft12\delta \tau)\right\}\gO_{x}(\tau)\\
\algebra{P}_{x}(\tau+\delta\tau)&=\algebra{P}_{x,\mu}(\tau+\ft12\delta\tau)
+\ft12\delta\tau\, \dot{\algebra{P}}_{x}(\tau+\delta\tau)\,.
\end{aligned}
\end{equation}

\renewcommand{\eprint}[1]{ \href{http://arxiv.org/abs/#1}{[arXiv:#1]}}

\bibliography{paper}


\end{document}